\documentclass[12pt]{article}
\pdfoutput=1
\usepackage{amsmath,amssymb,amsfonts,graphics,graphicx,amscd,amsfonts,epsf,epsfig,color}
\usepackage[hidelinks,pdftex]{hyperref}
\usepackage{xcolor, soul}
\allowdisplaybreaks[4]

\date{}
\begin{document}

\begin{flushright}
UTHEP-728
\end{flushright}

\vspace{0.1cm}

\begin{center}
  {\LARGE

Partial Deconfinement

  }
\end{center}
\vspace{0.1cm}
\vspace{0.1cm}

\begin{center}
Masanori Hanada,$^a$ 
Goro Ishiki,$^{n,t}$
and Hiromasa Watanabe$^t$
\end{center}

\begin{center}
$^a$
School of Physics and Astronomy, and STAG Research Centre\\
University of Southampton, Southampton, SO17 1BJ, UK\\
\vspace{0.2cm}
$^n$
Tomonaga Center for the History of the Universe, \\
University of Tsukuba, Tsukuba, Ibaraki 305-8571, Japan\\
\vspace{0.2cm}
$^t$
Graduate School of Pure and Applied Sciences\\
University of Tsukuba, Tsukuba, Ibaraki 305-8571, Japan

\end{center}

\vspace{1.5cm}

\begin{center}
  {\bf Abstract}
\end{center}
We argue that the confined and deconfined phases in gauge theories are connected by 
a {\it partially deconfined} phase (i.e. SU$(M)$ in SU$(N)$, where $M<N$, is deconfined), 
which can be stable or unstable depending on the details of the theory. 
When this phase is unstable, it is the gauge theory counterpart of the small black hole phase in the dual string theory. 
Partial deconfinement is closely related to the Gross-Witten-Wadia transition, 
and is likely to be relevant to the QCD phase transition.  

The mechanism of partial deconfinement is related to a generic property of a class of systems.  
As an instructive example, we demonstrate the similarity between the Yang-Mills theory/string theory and a mathematical model of the collective behavior of ants
[Beekman et al., Proceedings of the National Academy of Sciences, 2001]. 
By identifying the D-brane, open string and black hole with the ant, pheromone and ant trail, 
the dynamics of two systems closely resemble with each other, 
and qualitatively the same phase structures are obtained. 

\newpage

\tableofcontents
\section{Introduction}
\hspace{0.51cm}
A longstanding problem in quantum field theory is understanding the details of the finite temperature
deconfinement transition in QCD, and gauge theories more broadly.
For QCD, this transition can be probed experimentally in heavy ion collisions, and can play an important role in the physics of the early universe.
A central difficulty in theoretically studying the transition is the presence of strong interactions, rendering perturbation theory futile and requiring numerical techniques. 

Certain strongly coupled gauge theories can be described by weakly coupled gravity via holographic duality \cite{tHooft:1993dmi,Susskind:1994vu,Maldacena:1997re}. 
According to the duality, the deconfinement transition in the gauge theory is equivalent to the formation of a black hole.
The duality allows us to learn about the nature of the deconfinement transition from the perspective of gravity, 
and at the same time, the microscopic quantum gravitational aspects of the formation and evaporation of black hole are encoded in gauge theory. 

\begin{figure}[htbp]
 \begin{minipage}{0.32\hsize}
 \begin{center}
   \rotatebox{0}{
  \includegraphics[width=50mm]{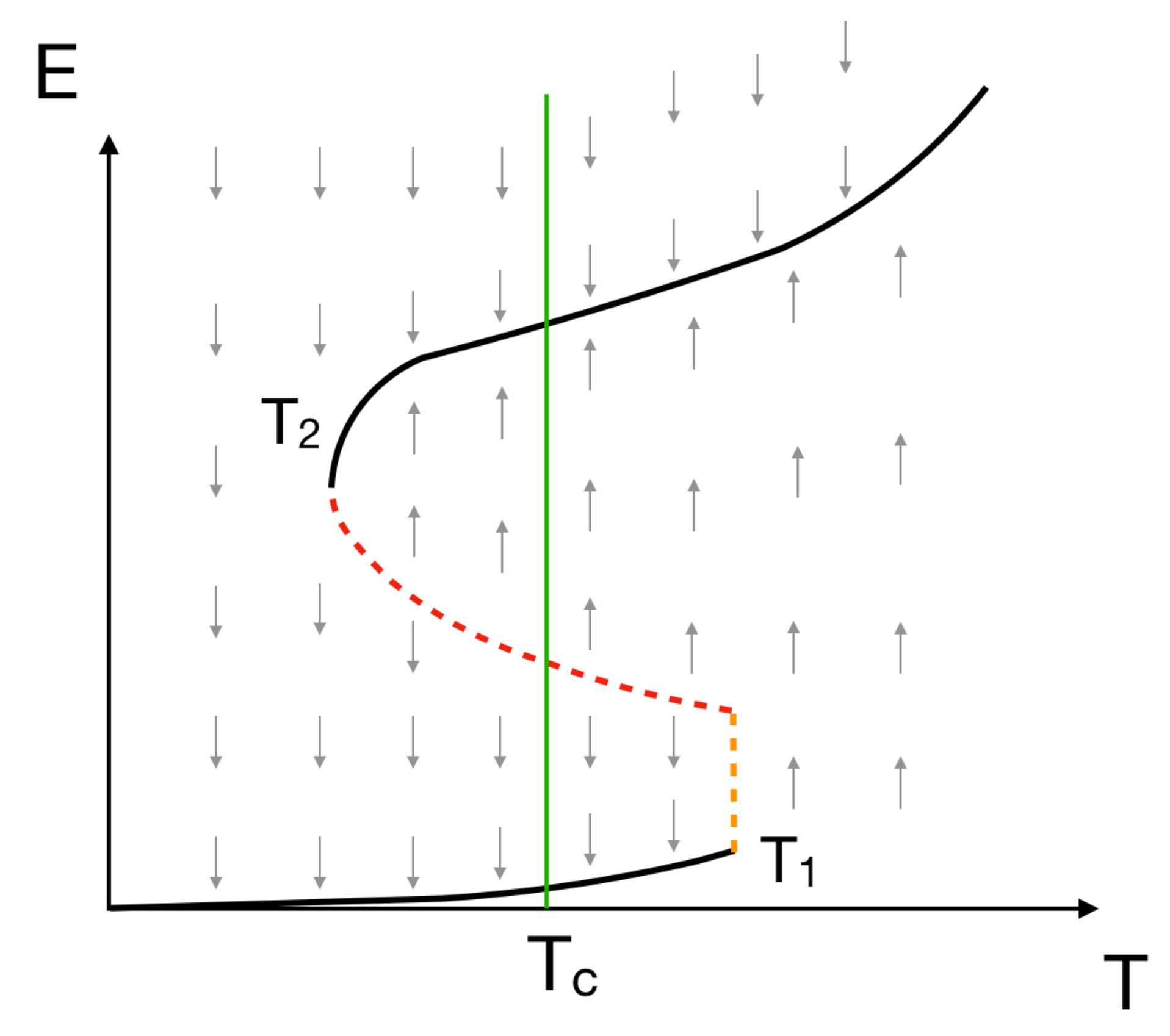}}
 \end{center}
 \end{minipage}
 \begin{minipage}{0.32\hsize}
 \begin{center}
   \rotatebox{0}{
  \includegraphics[width=50mm]{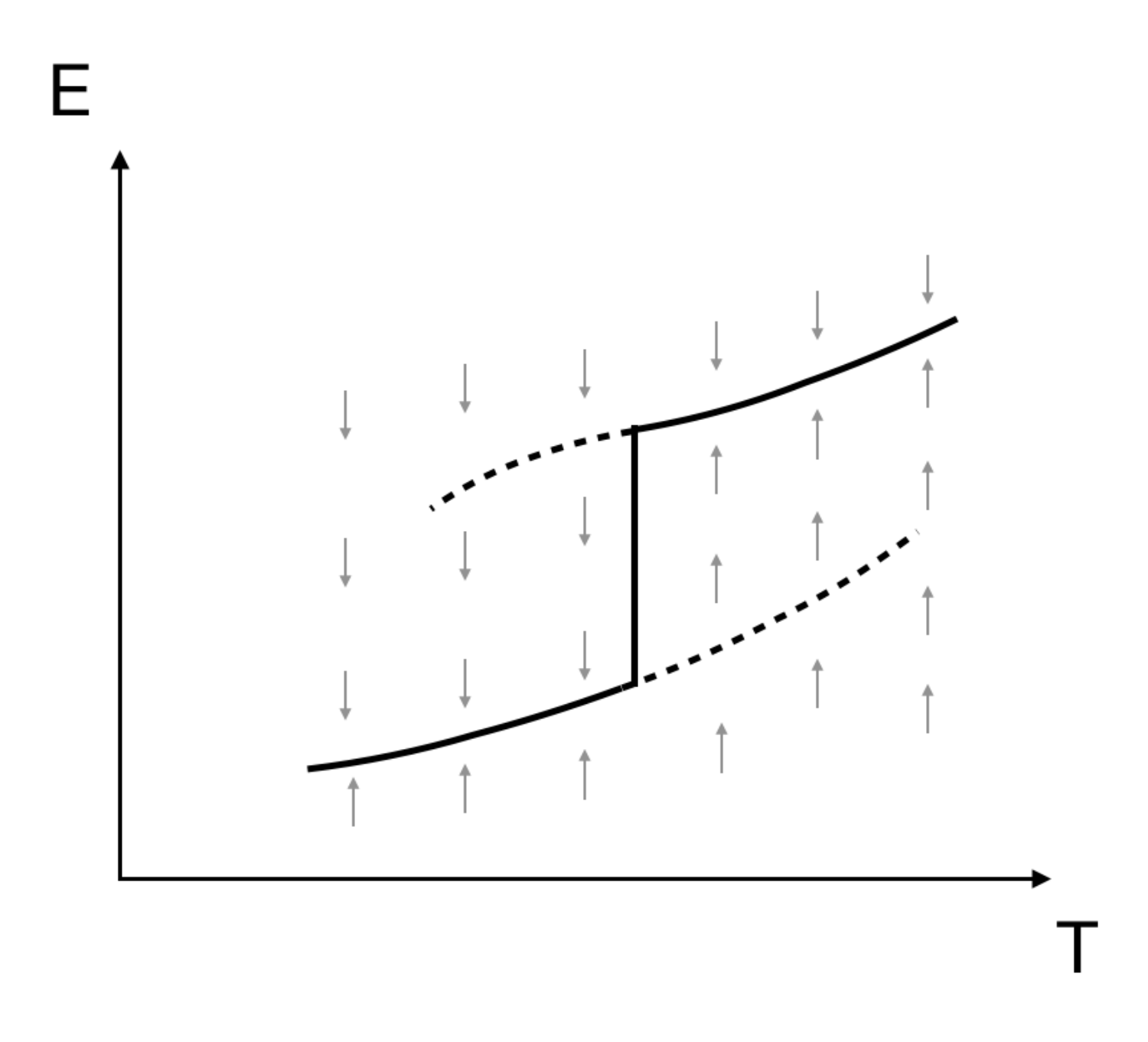}}
 \end{center}
 \end{minipage}
 \begin{minipage}{0.32\hsize}
 \begin{center}
   \rotatebox{0}{
  \includegraphics[width=50mm]{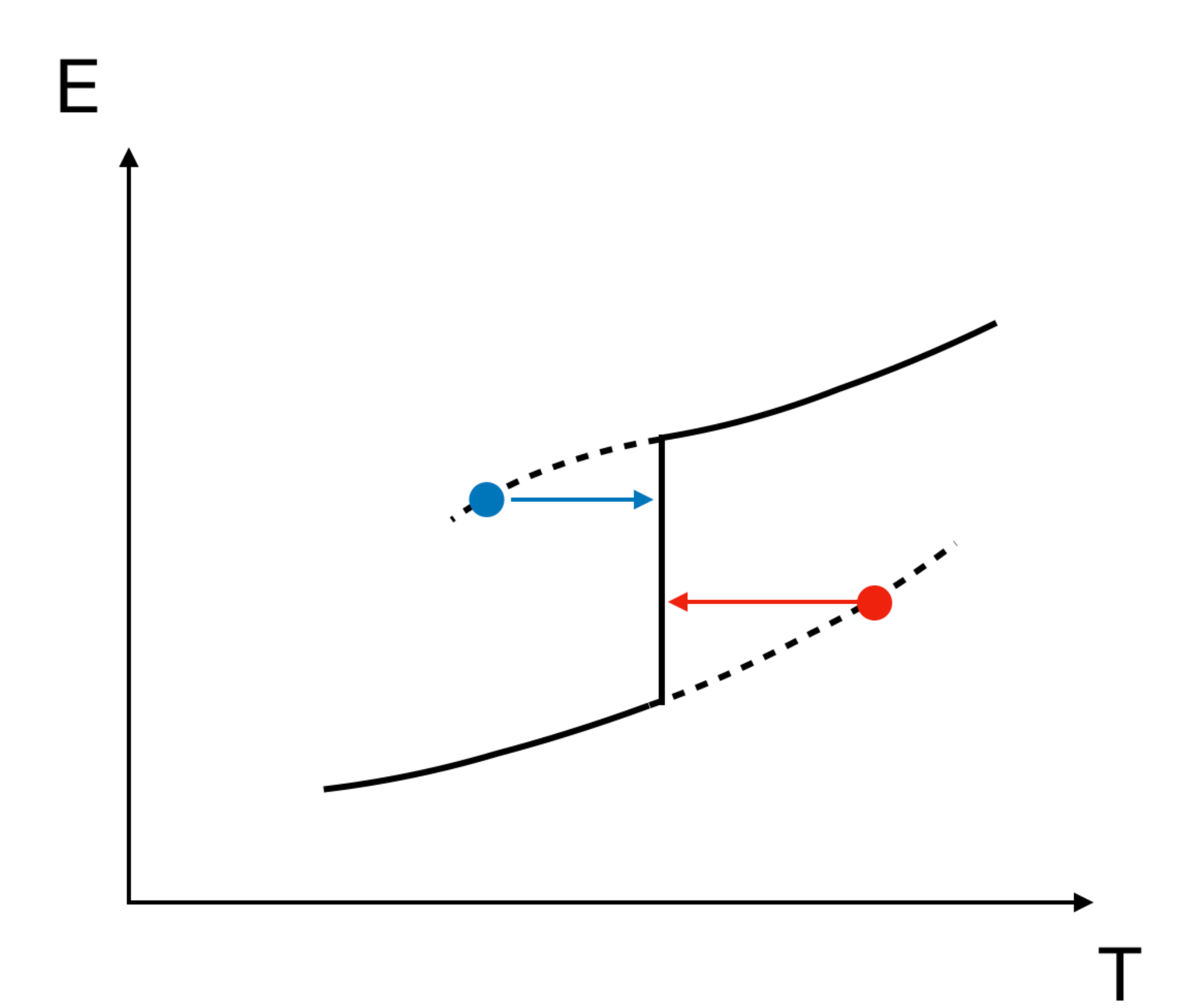}}
 \end{center}
 \end{minipage} 
 \caption{
[Left] The phase structure of 4D ${\cal N}=4$ SYM on S$^3$. The orange and red dashed lines denote the Hagedorn string phase 
 and small black hole phase, respectively, which are well-defined in the microcanonical theory. In the canonical treatment, the small black hole phase is the unstable saddle which is responsible for strong hysteresis. 
The green line marks the transition temperature in the canonical ensemble, $T_c$. 
Above $T_c$ the large black hole phase is favored, while below $T_c$ the graviton gas phase is favored.
[Center] A more common example of a first order transition, for example, between ice and liquid water. Small perturbations can destabilize the metastable states. 
At the critical temperature, there can be a mixture of two phases, in which ice and liquid water coexist.
[Right] Superheated ice (red point) and supercooled liquid water (blue point) are not stable even in the microcanonical ensemble. There is an instability toward the mixture of the ice and liquid water. 
 }\label{fig:color-vs-space}
\end{figure}

The most well-studied example of holographic duality is the one between 4d ${\cal N}=4$ super Yang-Mills on the three sphere 
and type IIB superstring theory on AdS$_5 \times$S$^5$ spacetime \cite{Maldacena:1997re,Witten:1998zw}. 
One can infer and study the microcanonical phase structure of the strongly coupled region of super Yang-Mills 
from the weakly coupled gravity dual, which is depicted in the left panel of Fig.~\ref{fig:color-vs-space}.
There are a large black hole phase (black solid line at large energy region), 
a small black hole phase (red dashed line), a Hagedorn string phase (orange dashed line), and a graviton gas phase (black solid line at low energy region)
\cite{Witten:1998zw,Aharony:2003sx}.  

The large black hole has a positive specific heat, i.e. energy increases with temperature, as $E\sim N^2 T^4$. 
On the other hand, the small black hole has a negative specific heat. 
When the black hole is much smaller than the curvature scale of AdS$_5 \times$S$^5$, 
it is approximately the same as the Schwarzschild black hole in ten spacetime dimensions, and $E\sim N^2 T^{-7}$. 
This small black hole phase is interesting, in part because it provides a microscopic description of black holes with negative specific heat, which is a proxy for evaporating black holes.
In terms of the canonical ensemble, the small black hole phase is described by an unstable saddle, in the following sense. 
Let us write the canonical partition function at temperature $T$ as 
\begin{eqnarray}
Z(T)
=
\int dE \,\Omega(E)\,e^{-E/T}
=
\int dE \,e^{-F(E,T)/T}, \label{eq:canonical-partition-function}
\end{eqnarray}
where $\Omega(E)=e^{S(E)}$ is the density of states, $S(E)$ is the microcanonical entropy, 
and $F(E,T)=E-TS$ is the free energy. The saddles of $F$ for fixed $T$ satisfy
\begin{eqnarray}
0
=
\frac{\partial F}{\partial E}
=
1-T\frac{dS}{dE}\,,  
\end{eqnarray}
where $(dS/dE)^{-1}$ is, by definition, the temperature in the microcanonical ensemble. 
Hence, the value of the energy $E(T)$ at the saddle, as a function of the temperature $T$, is the energy in the microcanonical ensemble. 
The stable saddles (local minima) correspond to the graviton gas and large black hole, whereas the unstable saddle (local maximum) 
corresponds to the small black hole.  In the canonical ensemble (i.e. for fixed $T$), if we drive the system away from the minimum, it rolls back to the same minimum.
The direction of the change, which is determined by the sign of $dS/dE$, is shown by gray arrows in the left panel of Fig.~\ref{fig:color-vs-space} above.  
In other words, the metastable phases (i.e. the large black hole phase at $T<T_c$ and the graviton gas phase at $T>T_c$) are stable against small perturbations.
This stability is unlike many standard examples of first order phase transitions, such as the transition between ice and liquid water.
For example, we can supercool liquid water below its freezing temperature, but ice appears as soon as a tiny perturbation is added.
In the case of ice and liquid water, as shown in the center panel of Fig.~\ref{fig:color-vs-space}, the mixture of two phases can exist at $T=T_c$ (the solid vertical line).
The direction toward equilibration is shown by gray arrows, and there is no unstable saddle.\footnote{
More precisely, although the unstable saddle could exist very close to the `metastable' state,
a tiny perturbation which does not depend on the volume of the system is sufficient to escape from the metastable state 
by going beyond the unstable saddle. 
In Yang-Mills, the size of the perturbation necessary for going beyond the unstable saddle increases with $N$. 
In this sense, the `metastable' phase in Yang-Mills is actually `stable' in the large-$N$ limit. 
}
Even in the microcanonical ensemble, as shown in the right panel of the figure, the supercooled and superheated phases are unstable towards the mixture of ice and liquid water.\footnote{
In the high energy theory/string theory community, the phrase `two-phase coexistence' is sometimes used sloppily, and has two different meanings:
(i) the mixture of two phases, such as liquid and solid phases, and (ii) existence of two (meta-)stable phases at the same temperature. 
In order to avoid the confusion, we will not use this terminology. 
}
Such instability is absent in the microcanonical treatment of the black hole and super-Yang-Mills.   

From the gravitational point of view, the phase structure in the left-panel of Fig.~\ref{fig:color-vs-space} is easy to understand \cite{Witten:1998zw,Aharony:2003sx}.
But by holographic duality, there is necessarily an alternative description of the phase structure in terms of the dual gauge theory.
While various features of the phase structure have familiar counterparts in gauge theories, the counterpart to the small black hole phase is unfamiliar,
and begs a gauge theory description.
Previously, it has been proposed \cite{Hanada:2016pwv} that the mechanism which we call {\it partial deconfinement} can naturally explain the gauge theory counterpart to the small black hole phase. 
Berenstein \cite{Berenstein:2018lrm} used a simple matrix model and combinatorial arguments to justify partial deconfinement. His arguments suggest that partial deconfinement is a generic feature of gauge theories, 
even without a gravity dual. 
In this paper, we leverage gauge theory calculations and numerics to provide compelling evidence in support of partial deconfinement, 
and give an intuitive explanation of the mechanism behind it. 
In particular, we will see that the distribution of Polyakov line phases contains rich information 
about the partially deconfined phase. 
Our results suggest that partial deconfinement is universal across a broad class of supersymmetric and non-supersymmetric gauge theories, and approximately applies to real-world QCD. 

This paper is organized as follows. 
In Sec.~\ref{sec:partial-deconf}, we introduce the notion of partial deconfinement. 
In Sec.~\ref{sec:ant-BH-correspondence}, 
we show that the partial deconfinement phase in Yang-Mills is well-captured by models of collective motion with positive feedback.  For concreteness, we consider a well-known model of ant trail formation.
Intuition from the ant model leads us to a unified perspective on the phase structures of various 
Yang-Mills theories with and without first order transitions.
In Sec.~\ref{sec:polyakov_line}, we provide quantitative analytic and numerical evidence of the partial deconfinement phase in various models.
In Sec.~\ref{sec:discussion}, we conclude with a discussion, 
including potential application to QCD.
We collect various details of our calculations and numerical simulations in the Appendices.

\section{Partial deconfinement}\label{sec:partial-deconf}
\hspace{0.51cm}
As mentioned in the introduction, 4d ${\cal N}=4$ SYM has a peculiar phase structure. 
Its peculiarities can be understood by examining the kinds of degrees of freedom which characterize various phases.
In the water/ice example two phases can occupy different regions in space at the freezing temperature, 
and such mixture of two phases connects the completely-liquid and completely-solid phases. 
We can supercool or superheat the system, but 
however large the volume of the full system, a perturbation localized in space can create a bubble of a more stable state, which then spreads to fill the entire volume.
In the 4d ${\cal N}=4$ SYM, it is natural to assume  
{\it two phases (confined and deconfined) can occupy different regions in color degrees of freedom} \cite{Hanada:2016pwv,Asplund:2008xd,Berenstein:2018lrm}, 
namely SU$(M)$ in SU$(N)$, where $M<N$, is deconfined. 
Let us call it {\it partial deconfinement}, as coined in Ref.~\cite{Berenstein:2018lrm}. 

In the language of D-branes and open strings, $M$ of $N$ D-branes are connected by open strings and 
form a bound state (black hole).
It can be regarded as `the mixture of two phases' from the D-brane point of view. 
The difference from the example of water is that the locations of D-branes is described by the internal degrees of freedom (color degrees of freedom) in QFT, rather than the spatial coordinate. 
Note that the interaction between D-branes can be highly nonlocal, in that all D-branes in the bound state interact with each other via open strings. This is in a stark contrast with the case of water. 

In some cases, partial deconfinement can lead to a negative specific heat
because {\it the number of unlocked degrees of freedom changes} \cite{Berkowitz:2016znt,Hanada:2016pwv}. 
In the partially deconfined phase, the number of unlocked degrees of freedom participating in the dynamics is proportional to $M^2$, rather than $N^2$.
(In terms of string theory, $M$ D-branes are forming a bound state, and open strings between them are excited.)
When the energy $E$ is increased, more strings can be excited, and hence $M$ increases with $E$. But then the energy per degree of freedom ($\simeq$ temperature), which is proportional to $E/M^2$, can increase or decrease, depending on the details of how $M$ depends on $E$. 
Thus negative specific heat can arise \cite{Berkowitz:2016znt,Berkowitz:2016muc}. 
This is in sharp contrast to the completely deconfined phase,
in which the number of degrees freedom is fixed and hence the specific heat has to be positive. 

Our general arguments so far do not establish exactly what theories have the negative specific heat. 
In Ref.~\cite{Hanada:2016pwv}, the properties of 4d ${\cal N}=4$ SYM has been used to explain 
$E/N^2\sim T^{-7}$ at low energies, which agrees with the equation of state of the ten-dimensional Schwarzschild black hole 
and hence is consistent with the AdS/CFT duality (see Appendix~\ref{sec:Hanada-Maltz}). 

As another set of examples, large-$N$ pure Yang-Mills in 3d and 4d flat space with large volume are known to have first order deconfinement transitions. 
Numerical simulations demonstrate the existence of strong hysteresis which resembles the left panel of Fig.~\ref{fig:color-vs-space} \cite{Liddle:2008kk,Lucini:2003zr}. 
This can also be understood as a phase separation in the color degrees of freedom 
due to the unstable saddle (partial deconfinement):
although spatially localized small bubble of a more stable state (the confined phase surrounded by the completely deconfined phase, or vice versa) 
can spread once it is created, 
the creation of the bubble itself can be suppressed at large $N$.
The natural scale of the bubble is $1/\Lambda_{\rm QCD}$, which is $N$-independent, and hence the size of the perturbation needed for 
the creation of the bubble increases with $N$. (Note that we are considering the 't Hooft large-$N$ limit,
and hence large $N$ is taken before large volume. If the large volume is taken first, 
then a large enough fluctuation which goes beyond the partially deconfined phase would be generated somewhere.
Note also that the same suppression can work for the spinodal decomposition as well.)
Hence, as long as we start with the confined or completely deconfined phase and dial the temperature, 
we expect strong hysteresis.

Note that the partial deconfinement requires $N\ge 3$. Therefore, the SU(2) YM should not have the first order transition --- and it is actually the case, as demonstrated by lattice Monte Calro simulations. 

In our language, the small black hole and Hagedorn string 
(the red and orange dashed lines in the left panel of Fig.~\ref{fig:color-vs-space}) are partially deconfined. 
Partial deconfinement implies that the distribution of the Polyakov line phase should behave as\footnote{We will show the heuristic derivation in in Sec.~\ref{sec:polyakov_line}. 
}
\begin{eqnarray}
\rho(\theta)
=
\frac{N-M}{N}\rho_{\rm confine}(\theta)
+
\frac{M}{N}\rho_{\rm deconfine}(\theta)
=
\frac{N-M}{N}\cdot\frac{1}{2\pi}
+
\frac{M}{N}\rho_{\rm deconfine}(\theta),  
\label{eq:partial-deconfinement}
\end{eqnarray}
where $\rho_{\rm deconfine}(\theta)$ is the distribution at the endpoint of the fully deconfined phase, $T=T_2$. 
At least in the examples we study in this paper, the completely deconfined phase is gapped
(i.~e. $\rho(\theta)=0$ at $\theta=\pm\pi$), and  
the partially deconfined phase is the non-uniform un-gapped phase.
We conjecture that this is true in general. 

In the past it has been argued that the gapped phase in 4d ${\cal N}=4$ SYM should be dual to the large black hole with positive specific heat \cite{Aharony:2003sx,AlvarezGaume:2006jg}. 
It is consistent with our observation, because we are relating the partially deconfined phase 
to the negative specific heat.

\section{Intuition from black hole/ant trail correspondence}\label{sec:ant-BH-correspondence}
\hspace{0.51cm}
The first order transition with an unstable saddle can be made transparent by considering a similar phase transition in the science of complex systems: 
the formation of the ant trail. 
This is one of the basic problems in the science of the collective animal behavior, 
and a concrete mathematical model is given and tested experimentally \cite{Beekman9703}.
More broadly, this is a common feature of systems with collective motion and positive feedback. 
A crucial feature is that when many ants follow the same trail, the strength with which other ants are drawn to follow to trail is enhanced. Likewise, D-branes forming a large bound state have enhanced interactions which D-branes not in the bound state. By identifying ants and D-branes in an appropriate manner, the formation of the ant trail can be identified with the formation of a black hole, and remarkably we can reproduce the essential features of the black hole.

Consider a colony of $N$ ants, where we take $N$ to be parametrically large. To simplify matters, we assume there is only one source of food, which is concentrated at some particular location. As ants walk away from the nest, some will find the food source and bring pieces back to the nest, leaving a pheromone trail along the way to direct other ants to the food source. Let $N_{\rm trail}$ be the number of ants rallying on a single trail between the nest and the food source. 
The strength with which ants not on the trail are drawn to the trail is proportional to the amount of secreted pheromones $pN_{\rm trail}$, where $p$ is the pheromone contribution from each ant on the trail.
Hence, the interaction between ants not on the trail, and the pheromones along the trail, is enhanced as $N_{\rm trail}$ becomes larger.
This is similar to the case of gauge theory: if $N_{\rm BH}$ D-branes form a bound state and one of the remaining $N-N_{\rm BH}$ D-branes
comes close by,\footnote{
Below we use $N_{\rm BH}$ instead of $M$, because it is identified with the number of D-branes forming the black hole. 
} there are $N_{\rm BH}$ different open strings which try to capture that D-brane.\footnote{
This mechanism is also known as the moduli trapping \cite{Kofman:2004yc} and is applied to the inflationary cosmology. 
}$^,$\footnote{
This can also be interpreted as the entropic force associated with the enhancement of the degrees of freedom 
from $N_{\rm BH}^2$ to $(N_{\rm BH}+1)^2$. 
The entropic force exists for any gauge theory, regardless of the existence 
of the dual gravity description.
} At higher temperatures, each open string mode 
can be more highly excited and thus can contribute more to the dynamics.
Therefore, higher temperatures
correspond to larger values of the pheromone parameter. 
Hence the basic correspondence is
\begin{eqnarray}
N_{\rm trail} & \longleftrightarrow & N_{\rm BH},\\
p & \longleftrightarrow & T. 
\end{eqnarray}

The phenomenological equation introduced in Ref.~\cite{Beekman9703} is\footnote{
In the original notation in Ref.~\cite{Beekman9703}, $N_{\rm trail}$ and $p$ are $x$ and $\beta$, respectively.  
}
\begin{eqnarray}
\frac{dN_{\rm trail}}{dt}
&=&
({\rm ants\ coming\ into\ the\ trail})
-
({\rm ants\ leaving\ the\ trail})
\nonumber\\
&=&
(\alpha+pN_{\rm trail})(N-N_{\rm trail})
-
\frac{sN_{\rm trail}}{s+N_{\rm trail}}.
\label{eq:ant-equation}
\end{eqnarray}
Here $\alpha$ describes the probability that each ant accidentally find the food source, 
and $s$ determines the rate that ants leave the trail.
The parameters $p$, $\alpha$ and $s$ depend on the species, geography around the colony, weather and so on. 
The size of the stationary ant trail can be obtained by solving the equation $\frac{dN_{\rm trail}}{dt}=0$. 

\begin{figure}[htbp]
 \begin{minipage}{0.32\hsize}
 \begin{center}
   \scalebox{1.3}{
  \includegraphics[width=50mm]{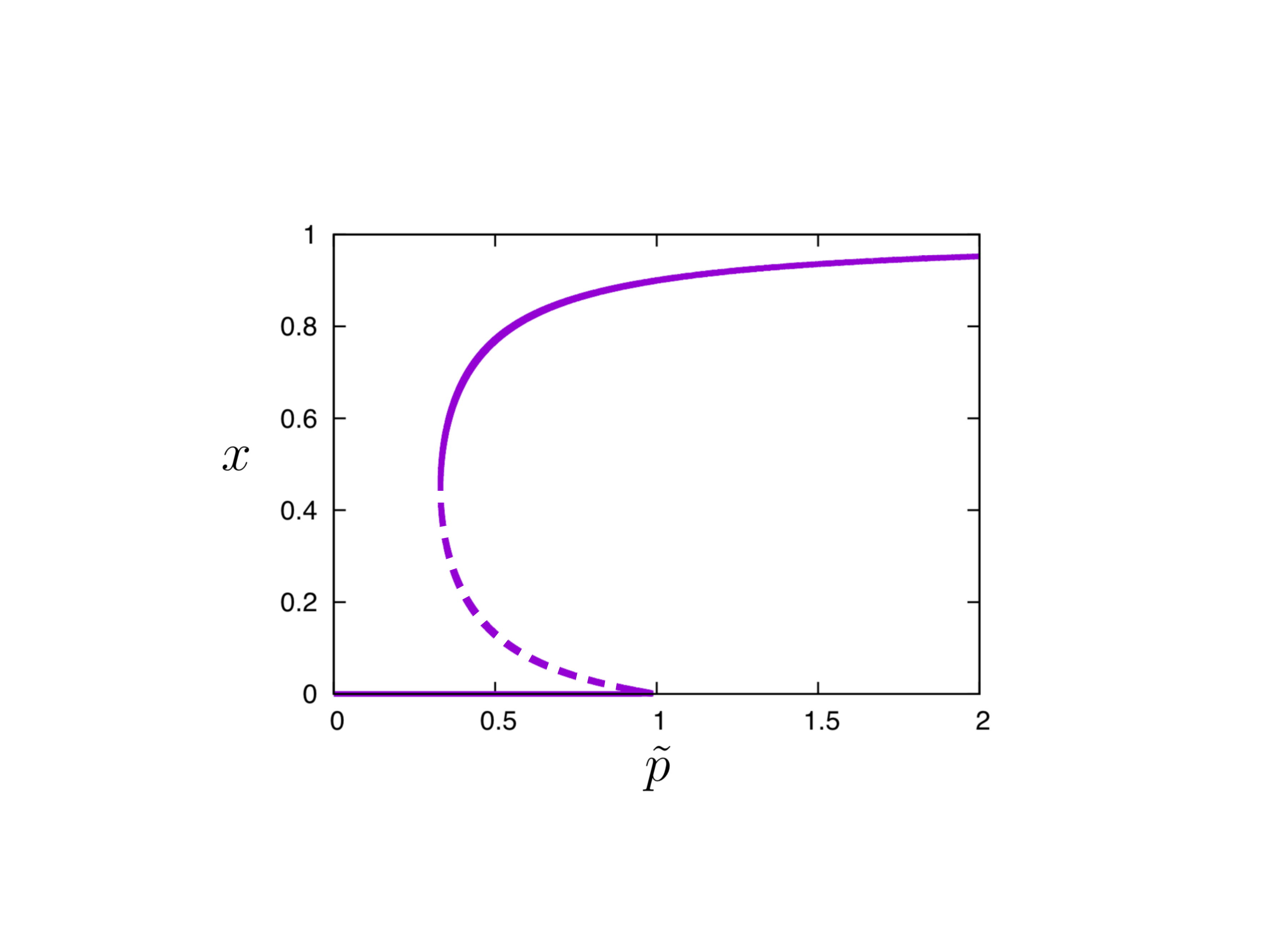}}
 \end{center}
 \end{minipage}
 \begin{minipage}{0.32\hsize}
 \begin{center}
   \scalebox{1.3}{
  \includegraphics[width=50mm]{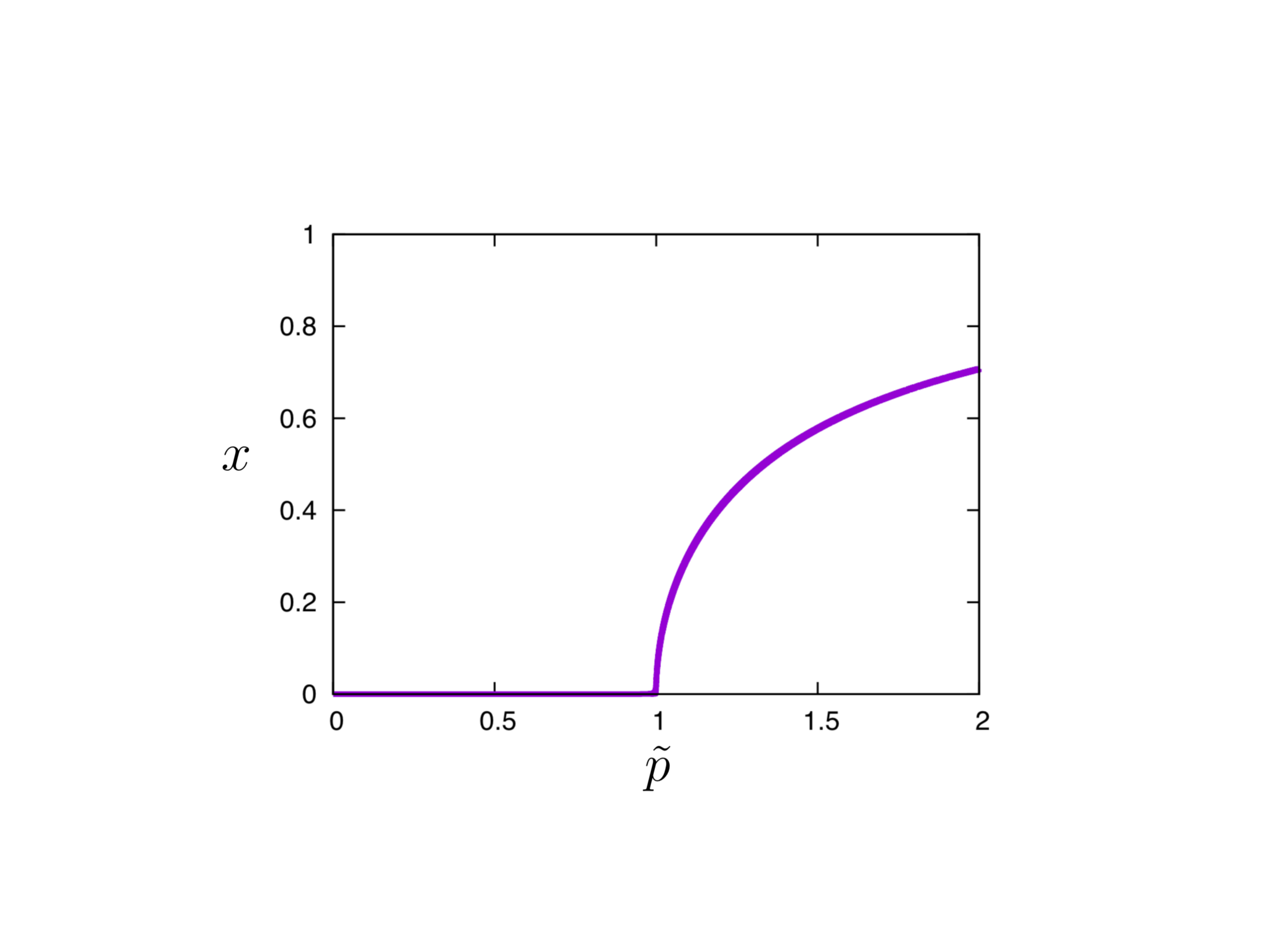}}
 \end{center}
 \end{minipage}
  \begin{minipage}{0.32\hsize}
 \begin{center}
   \scalebox{1.3}{
  \includegraphics[width=50mm]{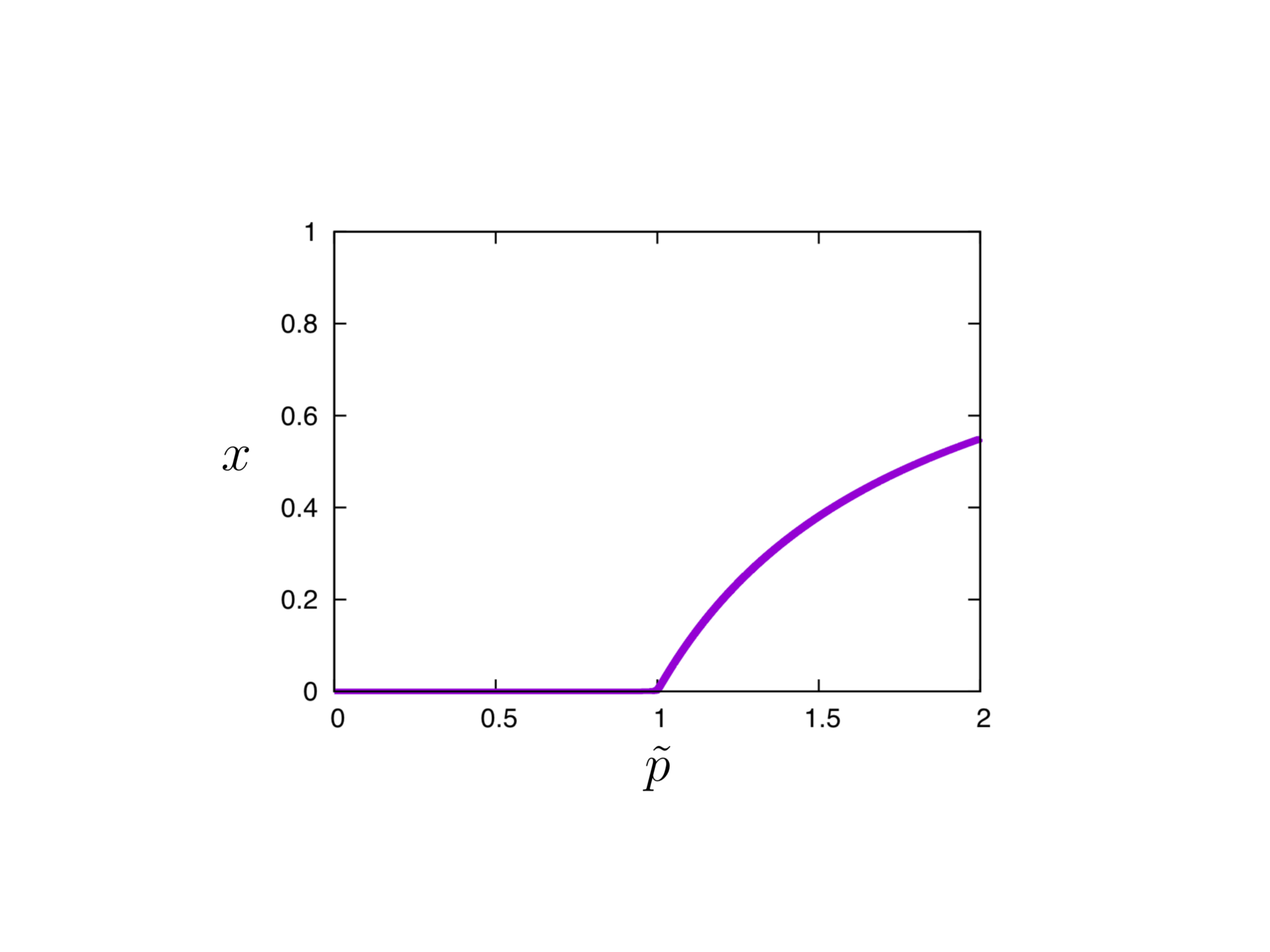}}
 \end{center}
 \end{minipage}
\caption{
$x\equiv\frac{N_{\rm trail}}{N}$ versus $\tilde{p}=Np$ in the ant trail model \eqref{eq:ant-equation}. 
$\alpha=\frac{1}{N}$, $\tilde{s}\equiv\frac{s}N=0.1$ (left), $1.0$ (center) and $5.0$ (right), 
$N=10^5$.  
 }\label{fig:Ntrail-vs-p}
\end{figure}

In the original treatment \cite{Beekman9703}, $p$, $\alpha$ and $s$ are fixed and $N_{\rm trail}$ has been calculated 
as the function of $N$. 
For our purposes, in order to make contact with black holes and gauge theory, 
we fix $\alpha$, $s$ and $N$, and calculate $N_{\rm trail}$ as the function of $p$. 
As discussed in Appendix~\ref{sec:ant-large-N}, the interesting large-$N$ limit of the ant model 
analogous to black hole is 
$\alpha\sim N^{-1}$, $p\sim N^{-1}$, $s\sim N^1$. 
In Fig.~\ref{fig:Ntrail-vs-p}, we show the plot of $x\equiv\frac{N_{\rm trail}}{N}$ versus $\tilde{p}\equiv Np$ 
for $\alpha=\frac{1}{N}$, $\tilde{s}\equiv\frac{s}N=0.1,1.0, 5.0$, and $N=10^5$. 
(See Appendix~\ref{sec:ant-large-N} for the analytic argument and `physical' intuition.) 
The saddles (the solutions of $\frac{dN_{\rm trail}}{dt}=0$) appear when the inflow and outflow of the ants
(the first and second terms of \eqref{eq:ant-equation}) balance. 
There are two cases, with and without the unstable saddle (dashed line):
\begin{itemize}
\item
If the inflow/outflow wins when $N_{\rm trail}$ is varied slightly upwards/downwards from the saddle, 
the saddle is unstable (dashed line in the left panel of Fig.~\ref{fig:Ntrail-vs-p}). 
It happens at $\tilde{s}<1$.

\item
If the inflow/outflow wins when $N_{\rm trail}$ is varied slightly downwards/upwards from the saddle, 
the saddle is stable (solid line). 
At $\tilde{s}>1$, only the stable saddle exists.  

\item
The boundary of these two behaviors can be seen at $\tilde{s}=1$ (the center panel of Fig.~\ref{fig:Ntrail-vs-p}). 
The curve $x = x(\tilde{p})$ has infinite slope at $\tilde{p}=1$ in the large-$N$ limit. 

\end{itemize}

The crucial feature responsible for the emergence of the unstable saddle is the positive feedback: 
near the unstable saddle, if $N_{\rm trail}$ increases/decreases a little bit, then more ants join/leave the trail 
and $N_{\rm trail}$ increases/decreases even more. 
The value of $N_{\rm trail}$ at the unstable saddle decreases as $\tilde{p}$ is increased
--- negative specific heat, by interpreting larger $N_{\rm trail}$ corresponds to 
larger energy 
--- because with larger value of $\tilde{p}$ strong enough total pheromone can be obtained 
with smaller value of $N_{\rm trail}$.

\begin{figure}[htbp]
 \begin{minipage}{0.32\hsize}
 \begin{center}
   \rotatebox{0}{
  \includegraphics[width=40mm]{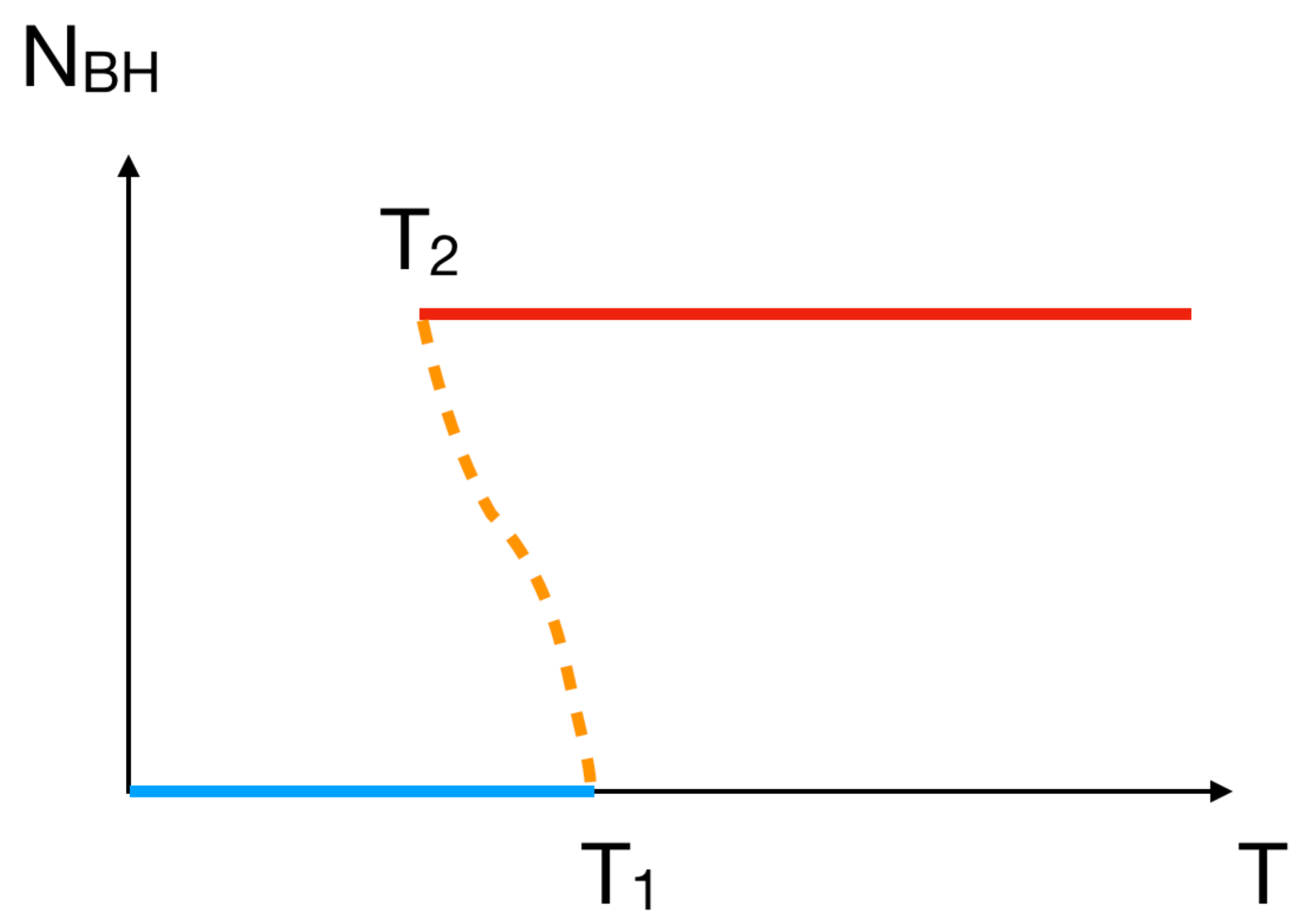}}
 \end{center}
 \end{minipage}
 \begin{minipage}{0.32\hsize}
 \begin{center}
   \rotatebox{0}{
  \includegraphics[width=40mm]{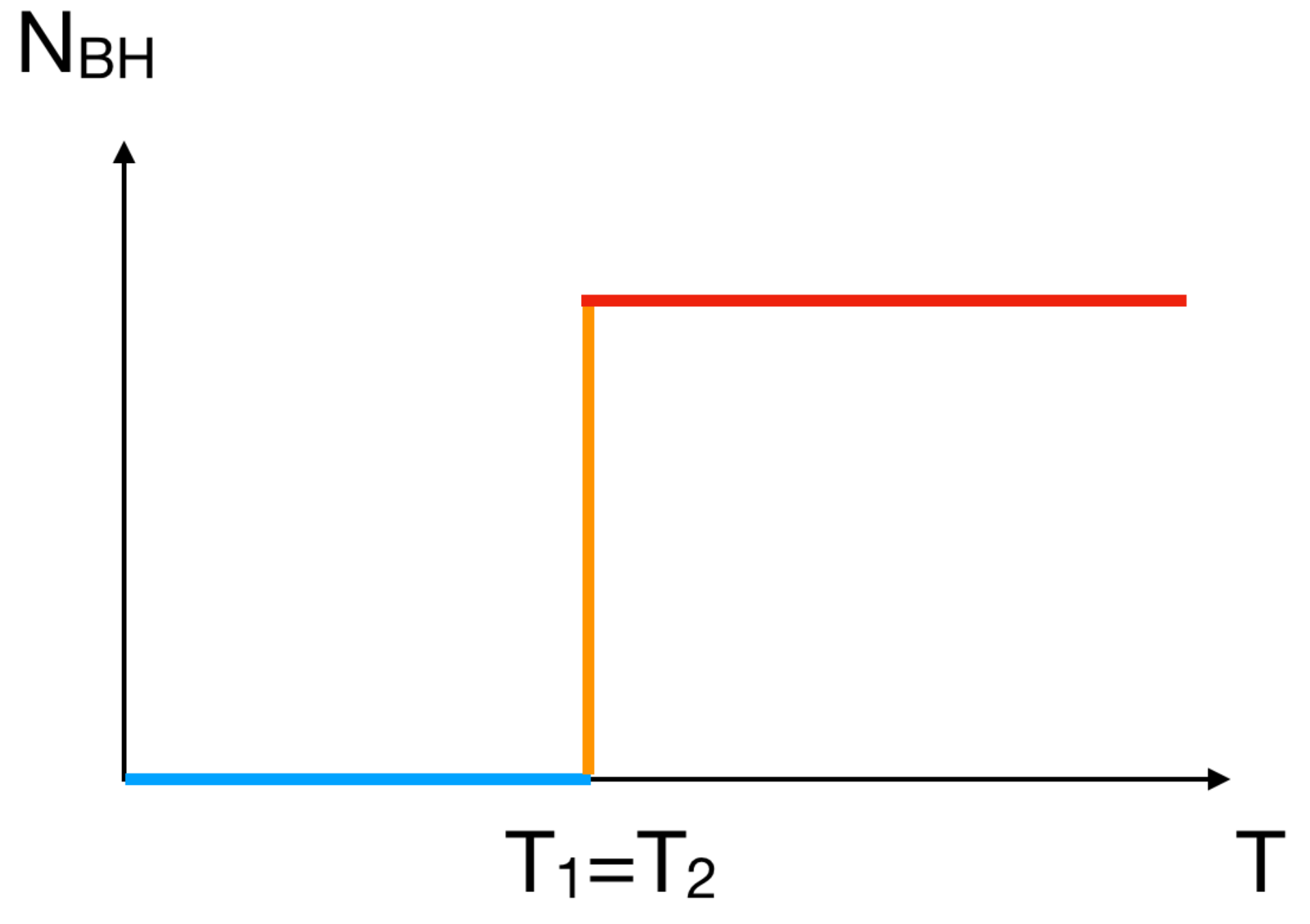}}
 \end{center}
\end{minipage}
\begin{minipage}{0.32\hsize}
\begin{center}
   \rotatebox{0}{
  \includegraphics[width=40mm]{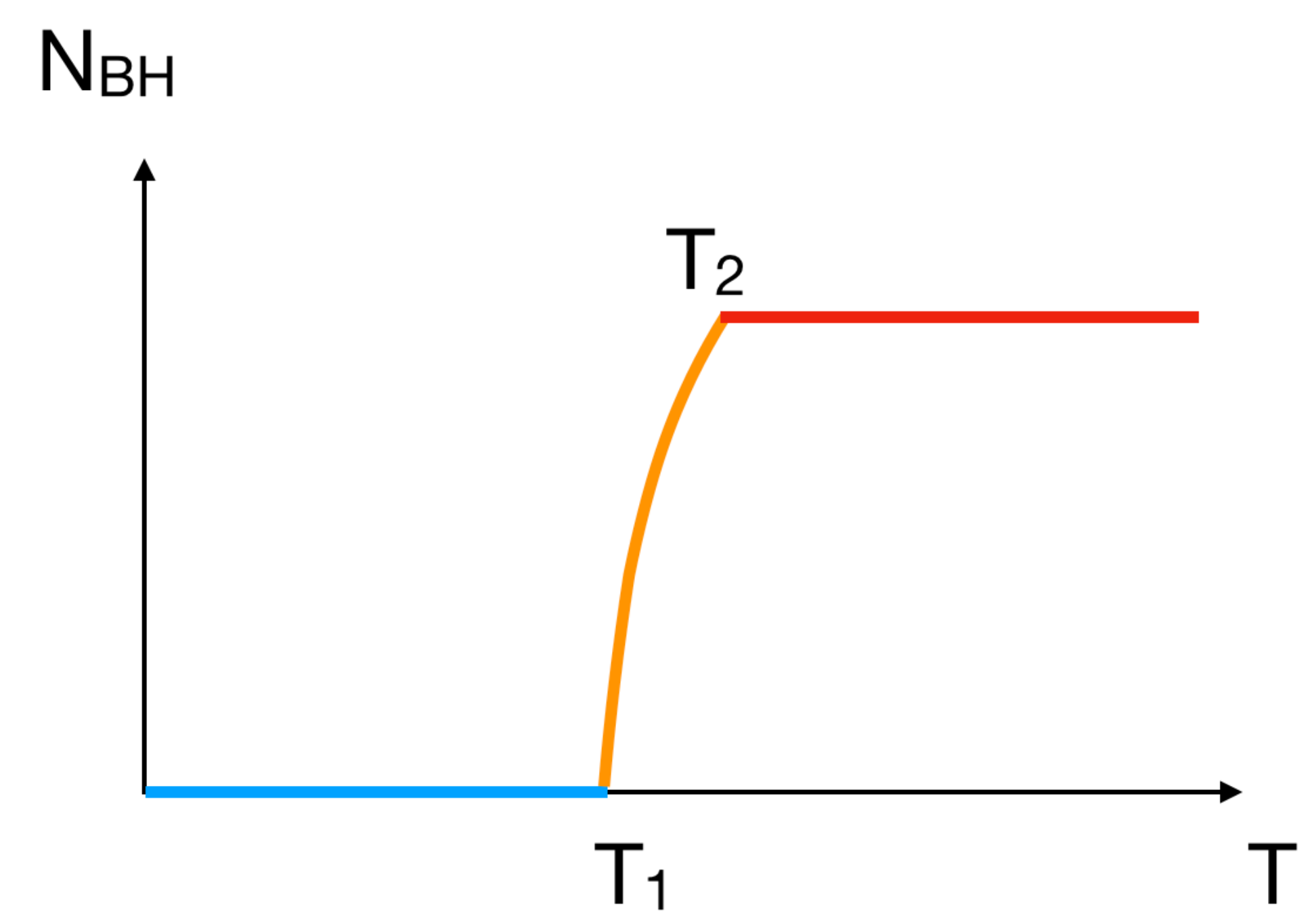}}
\end{center}
\end{minipage}
\caption{
Cartoon pictures of $T$ vs $N_{\rm BH}$ in gauge theory. The blue and red lines are confined and completely deconfined phases, 
where $N_{\rm BH}\sim O(N^0)$ and $N_{\rm BH}=N$, respectively. Orange lines indicate the partially deconfined phase. 
 }\label{fig:Nbh-vs-T}
\end{figure}

\begin{figure}[htbp]
 \begin{minipage}{0.32\hsize}
 \begin{center}
   \rotatebox{0}{
  \includegraphics[width=40mm]{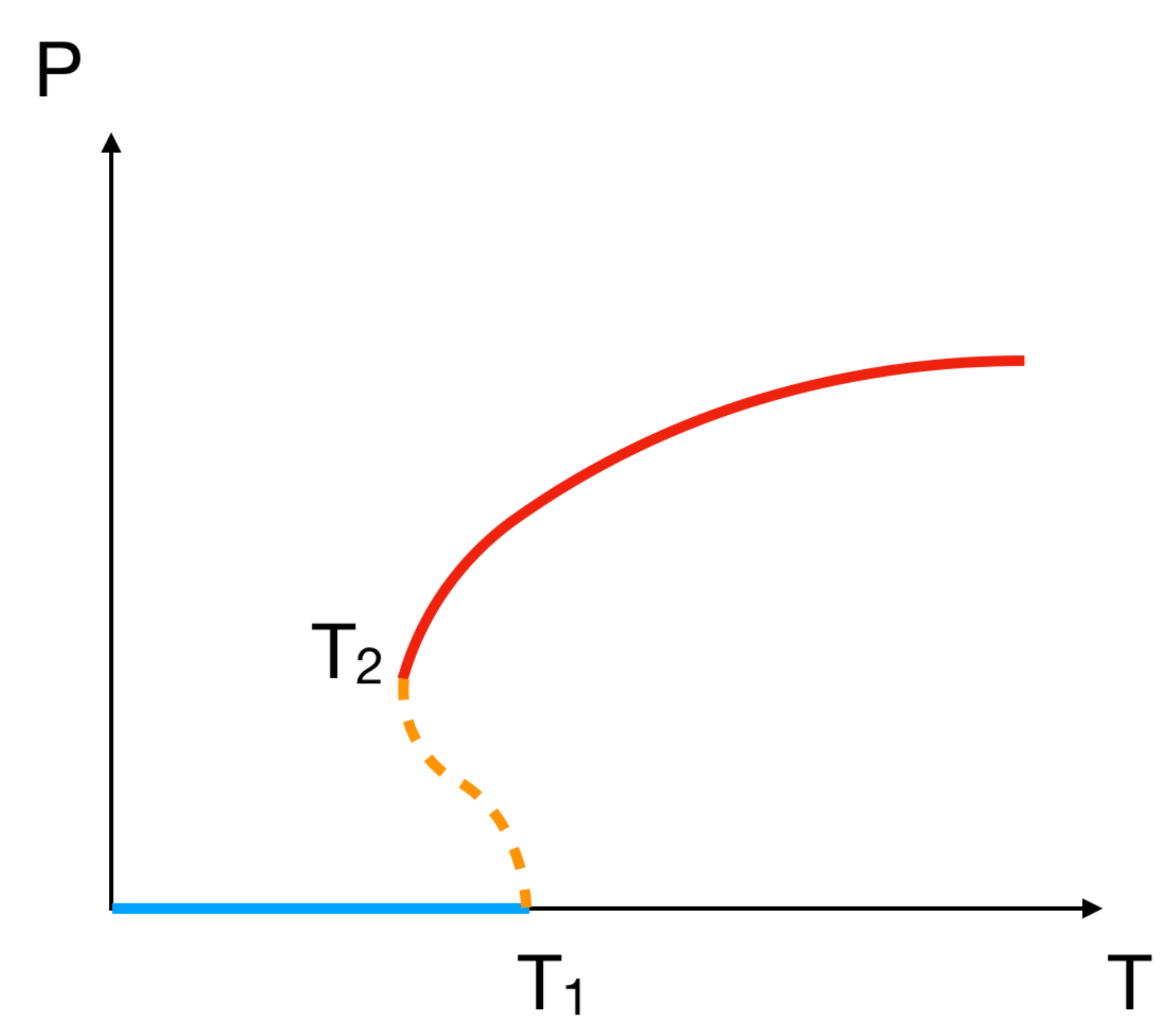}}
 \end{center}
 \end{minipage}
 \begin{minipage}{0.32\hsize}
 \begin{center}
   \rotatebox{0}{
  \includegraphics[width=40mm]{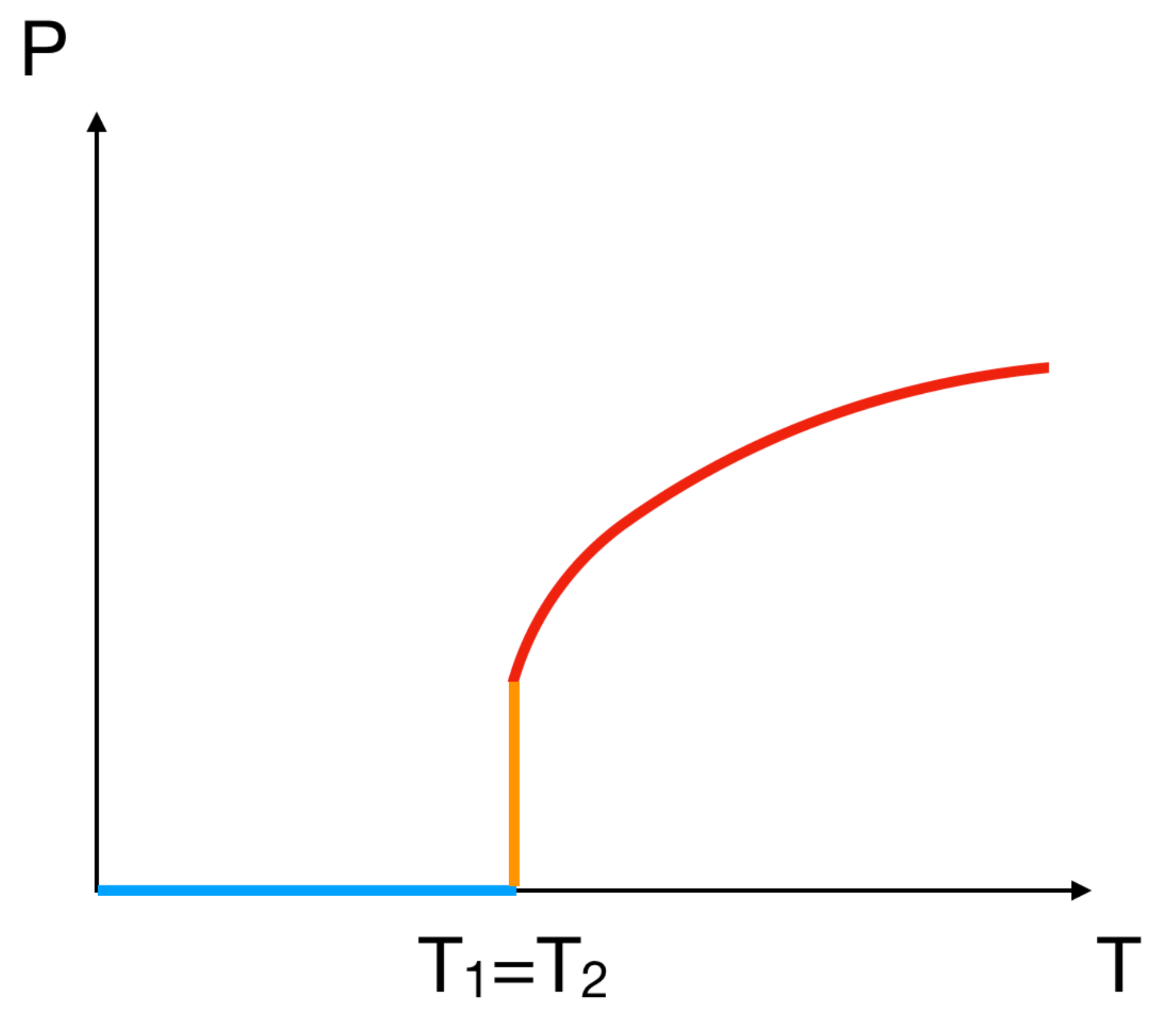}}
 \end{center}
 \end{minipage}
  \begin{minipage}{0.32\hsize}
 \begin{center}
   \rotatebox{0}{
  \includegraphics[width=40mm]{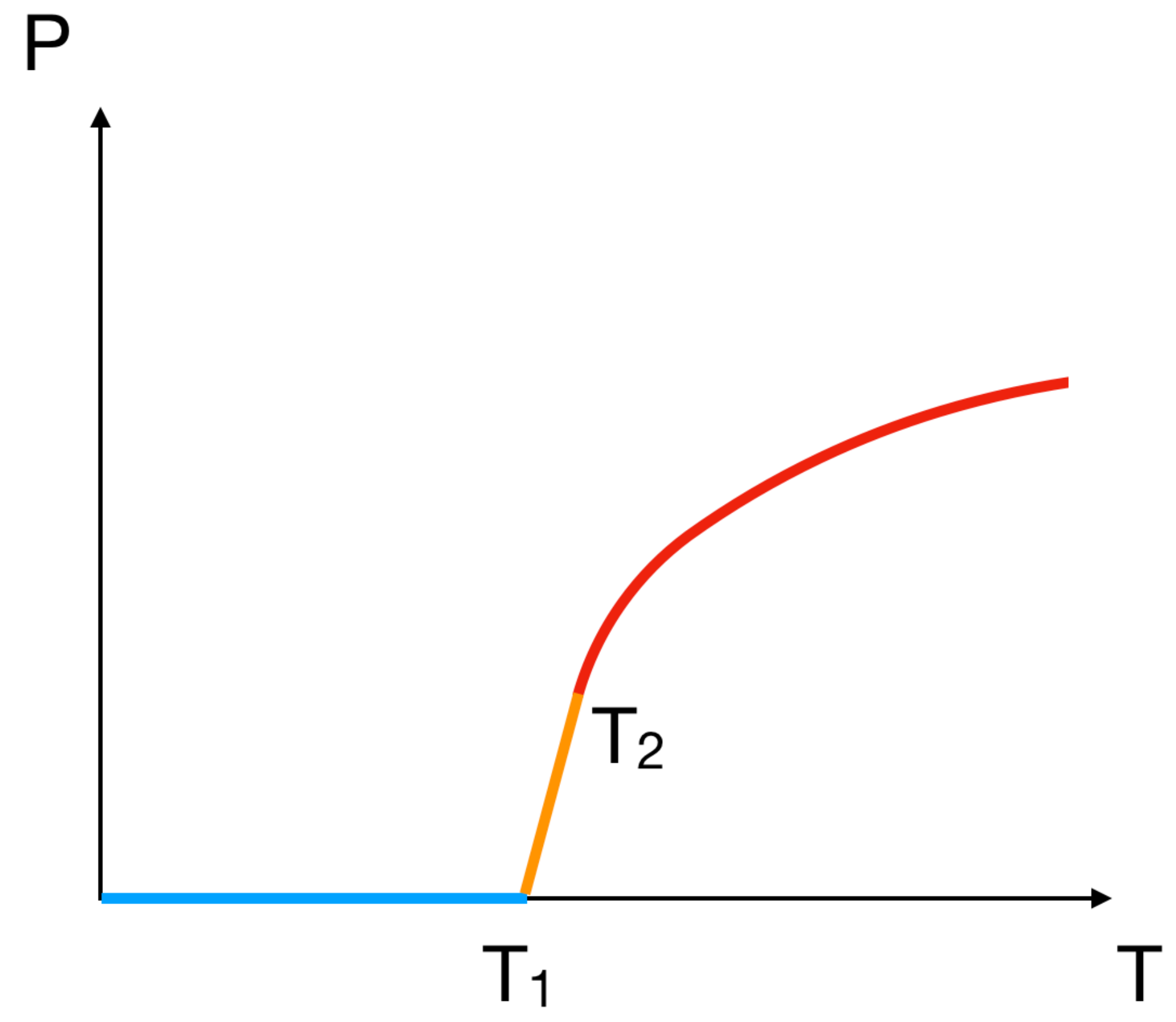}}
 \end{center}
 \end{minipage}
 \caption{
 Cartoon pictures of the Polyakov loop $P$ as a function of temperature. 
 The blue, orange and red lines represent the confined, partially deconfined and completely deconfined phases.  
 Similar curves are obtained by using $E/N^2$ as the vertical axis as well. 
 }\label{fig:bosonic-BMN}
\end{figure}

Essentially the same mechanism exists in gauge theory. In terms of strings and D-branes:
as $N_{\rm BH}$ grows, external D-branes are attracted more strongly (inflow); on the other hand, as $N_{\rm BH}$ grows, 
it costs more energy, and hence the Boltzmann factor pushes $N_{\rm BH}$ down (outflow); 
the saddle appears when these two effects balance. 
Again, whether the saddle is stable or unstable depends on the details of the dynamics. 
In Fig.~\ref{fig:Nbh-vs-T} and Fig.~\ref{fig:bosonic-BMN}, we show three possibilities:
\begin{itemize}
\item
If the inflow/outflow wins when $N_{\rm BH}$ is varied slightly upwards/downwards from the saddle, 
the saddle is unstable (the dashed line in the left panels of Fig.~\ref{fig:Nbh-vs-T} and Fig.~\ref{fig:bosonic-BMN}). 
In order for this to happen, the strings have to bind D-branes tighter, so the strong coupling dynamics is needed. 
Indeed, this is the case for the strongly coupled region of 4d ${\cal N}=4$ SYM. 

\item
If the inflow/outflow always balances, no particular value of $N_{\rm BH}$ is favored (the vertical solid line 
in the center panels of Fig.~\ref{fig:Nbh-vs-T} and Fig.~\ref{fig:bosonic-BMN}). 
This is the case for the weakly coupled region of 4d ${\cal N}=4$ SYM. 

\item
If the inflow/outflow wins when $N_{\rm BH}$ is varied slightly downwards/upwards from the saddle, 
the saddle is stable. In the right panels of Fig.~\ref{fig:Nbh-vs-T} and Fig.~\ref{fig:bosonic-BMN},  
only stable saddles exist. 
In 4d ${\cal N}=4$ SYM, there is no counterpart of the right panels. 

\end{itemize}

The unstable saddle exhibits the negative specific heat because at higher temperature 
each open string mode can be excited more 
(each ant contributes to more pheromone) and strong enough attraction (strong enough pheromone) resisting the emission of D-branes (outflow of the ants) can be achieved with smaller value of $N_{\rm BH}$. 

Despite the similarity in the dynamics and qualitative aspects of the phase structures, there is a difference as well: $N_{\rm trail}/N$ can be one only at $\tilde{p}=\infty$, 
namely the `complete deconfinement' is missing in the ant theory. 
In Appendix~\ref{sec:modified-ant-equation}, we show a similar model (Ant-Man model) which does capture the complete deconfinement.

Because it is not easy to see the value of $N_{\rm BH}$ manifestly in gauge theory,
we need more evidence to justify the picture shown in Fig.~\ref{fig:Nbh-vs-T}. 
Later in this paper, we will provide the evidence. Meanwhile, let's focus on Fig.~\ref{fig:bosonic-BMN} first, because is rather well-established. 
4d ${\cal N}=4$ SYM on S$^3$ has the same pattern as Fig.~\ref{fig:bosonic-BMN}, when the coupling constant $\lambda=g_{\rm YM}^2N$ is varied: 
as the coupling constant becomes smaller, the hysteresis becomes weaker ($T_2-T_1$ becomes smaller) and 
phase structure becomes like the center of Fig.~\ref{fig:bosonic-BMN} ($T_1=T_2$)
\cite{Aharony:2003sx,Sundborg:1999ue}. 
There are plenty of other examples which exhibit this pattern shown in Fig.~\ref{fig:bosonic-BMN}, including:
\begin{itemize}
\item
The real-world QCD has the rapid cross-over which resembles the right panel. 
By dialing the mass of the quark to sufficiently large values, the first order transition like the left panel can be realized. 

\item
We can also consider QCD at finite baryon chemical potential. It is widely believed that the first order transition shows up at sufficiently large baryon chemical potential.\footnote{
Usually the chiral transition, which can be different from deconfinement in general, is considered. See Sec.~\ref{sec:discussion} regarding this point. 
}

\item
The bosonic version of the plane wave matrix model, which we will study in detail in  Sec.~\ref{sec:BMN}. 

\item
In certain super Yang-Mills theories, 
the `deconfinement' along the spatial circle is related to the black hole/black string topology change. 
In this case the existence of the unstable saddle (non-uniform black string) can be shown from the gravity side. 
We will discuss the detail in Sec.~\ref{sec:BH-BS}. 

\end{itemize}

Based on these examples and the arguments in this section pertaining to collective motion with positive feedback,
we propose that the partial deconfinement is a generic feature of gauge theory. 
In Sec.~\ref{sec:polyakov_line}, we provide quantitative evidence based on the distribution of the Polyakov line phases.

\section{Quantitative tests with Polyakov line phases}\label{sec:polyakov_line}
\hspace{0.51cm}
In this section we show the evidence of the partial deconfinement in various settings,
both at weak coupling and strong coupling, 
based on the distribution of the Polyakov line phases. 
The relation \eqref{eq:partial-deconfinement} plays the central role. 
We find the universal behavior regardless of the choice of the theory.

Firstly we give a heuristic derivation of \eqref{eq:partial-deconfinement}.  
In the picture of the partial deconfinement, 
$T=T_2$ is the lowest possible temperature in the SU$(N)$ theory with the coupling constant $g_{\rm YM}^2$
at which all the D-branes can form a single bound state \cite{Hanada:2016pwv}.
Namely, if the temperature is below $T_2$, some D-branes are emitted from the bound state.  
Next let us consider a bound state consisting of $N_{\rm BH}$ of $N$ D-branes, 
which describes the partially deconfined phase. 
We assume that the dynamics of the bound state can be described by SU$(N_{\rm BH})$ theory, 
neglecting $N-N_{\rm BH}$ D-branes outside the bound state.  
Then, the temperature of the bound state is identified with 
the lowest possible temperature in the SU$(N_{\rm BH})$ theory with the coupling constant $g_{\rm YM}^2$ 
for which $N_{\rm BH}$ D-branes can form a single bound state \cite{Hanada:2016pwv}. 
Hence $N_{\rm BH}$ of the $N$ Polyakov line phases are distributed as $\rho_{\rm deconfine}(\theta)$, 
while the rest follow $\rho_{\rm confine}(\theta)=\frac{1}{2\pi}$. 

An implicit assumption here is that $\rho_{\rm deconfine}(\theta)$ does not depend on the effective 't Hooft coupling 
$g_{\rm YM}^2M$.
This assumption is satisfied for the examples we consider below: 
in 4d theories at weak coupling, the assumption can be confirmed by explicit calculations; 
in theories in less than four spacetime dimensions the 't Hooft coupling is dimensionful and hence it simply sets the 
unit of the energy scale.
(In case this assumption is not correct, we need to take into account the coupling dependence of $\rho_{\rm deconfine}(\theta)$.)

In all the cases we will study below, the transition between the partially and completely deconfined phases 
is the Gross-Witten-Wadia transition \cite{Gross:1980he,Wadia:2012fr,Wadia:1980cp}. 
We conjecture that more broadly, the partial-to-full deconfinement transition is universally described by 
Gross-Witten-Wadia. 
\subsection{4d ${\cal N}=4$ SYM on S$^3$}\label{sec:4dSYM_S3}
\hspace{0.51cm}
We begin with 4d ${\cal N}=4$ SYM on S$^3$. 
In the weak coupling limit, the phase distribution along the vertical orange line (i.e. `Hagedorn string' \cite{Hagedorn:1965st}) has been obtained analytically. 
The result is \cite{Sundborg:1999ue,Aharony:2003sx}
\begin{eqnarray}
\rho(\theta)
=
\frac{1}{2\pi}\left(
1+\frac{2}{\kappa}\cos\theta
\right),  \label{eq:phase_partial_deconfinement}
\end{eqnarray}
where $\kappa=2$ and $\kappa=\infty$ correspond to the intersection of the orange line
with the red and blue lines. 
If the orange line were not exactly vertical, these intersections would occur at the temperatures $T=T_2$ and $T=T_1$. 
One obtains a phase distribution of the same form as ~\eqref{eq:phase_partial_deconfinement}
for the unstable saddle which appears at finite coupling, as long as the coupling is sufficiently small.\footnote{
In Ref.~\cite{Aharony:2003sx}, the partition function in the free limit is expressed by using $u_n=\int d\theta \rho(\theta)e^{in\theta}$, 
as $Z=\int du_nd\bar{u}_n e^{-N^2\sum_{n\ge 1}c_n|u_n|^2}$. 
At the critical temperature, $c_n>0$ for $n\ge 2$, $c_1=0$, and hence $u_n=0$ for $n\ge 2$ and 
$u_1$ can take any value between 0 and 1, which leads to \eqref{eq:phase_partial_deconfinement}. 
At weak but finite coupling, 
the values of $c_n$ change, and higher order terms such as $|u_n|^4$ can appear. 
Still, $c_n>0$ holds for $n\ge 2$, and hence $u_n=0$ for $n\ge 2$ regardless of the detail. 
Therefore, \eqref{eq:phase_partial_deconfinement} also holds along the unstable saddle. 
}
We can rewrite $\rho(\theta)$ as 
\begin{eqnarray}
\frac{1}{2\pi}\left(
1+\frac{2}{\kappa}\cos\theta
\right) 
=
\left(
1
-
\frac{2}{\kappa}
\right)
\cdot
\frac{1}{2\pi}
+
\frac{2}{\kappa}
\cdot
\frac{1}{2\pi}\left(
1+\cos\theta
\right) 
\end{eqnarray}
and by identifying
\begin{eqnarray}
\frac{M}{N}
=
\frac{2}{\kappa}, 
\label{eq:partial-deconf-GWW}
\end{eqnarray}
we obtain \eqref{eq:partial-deconfinement}, 
where $\rho_{\rm deconfinement}(\theta)$ is the $\rho(\theta)$ at $\kappa=2$:
\begin{eqnarray}
\rho_{\rm deconfinement}(\theta)
=
\frac{1}{2\pi}\left(
1+\cos\theta
\right).  
\end{eqnarray}
The transition at $\kappa=2$ is the so-called Gross-Witten-Wadia transition. 

Therefore it is consistent with the partial deconfinement: the Hagedorn string can be identified with the partially deconfined phase.  
This identification, combined with the interpretation of black hole as long string \cite{sundborg1988strings,Susskind:1993ws,Horowitz:1996nw,Sundborg:1999ue,Aharony:2003sx,Hanada:2014noa}, 
naturally suggests that the unstable saddle (10d Schwarzschild black hole) at strong coupling is also in the partially deconfined phase. 

As discussed in Ref.~\cite{Hanada:2016pwv}, the partial deconfinement can explain the equation of state of the small black hole, $E\sim N^2T^{-7}$, 
and also other nontrivial features shown in the left panel of Fig.~\ref{fig:color-vs-space}. 
To make the paper self contained, we explain more details in Appendix~\ref{sec:Hanada-Maltz}. 

\subsection{Other 4d theories on S$^3$}
\hspace{0.51cm}
As discussed in Ref.~\cite{Aharony:2003sx}, 4d theories on the small three sphere can also behave
like the right panel of  Fig.~\ref{fig:bosonic-BMN}, depending on the details of the theories.  
The transitions are of the second order and third order at $T=T_1$ and $T=T_2$, respectively. 
Again, the latter is the Gross-Witten-Wadia transition. 
The Polyakov line phase distribution becomes as follows:
\begin{eqnarray}
\rho(\theta)
=
\left\{
\begin{array}{cc}
\frac{1}{2\pi} & (T\le T_1)\\
\frac{1}{2\pi}\left(
1+\frac{2}{\kappa}\cos\theta
\right) & (T_1<T<T_2)\\
\frac{2}{\pi\kappa}\cos\frac{\theta}{2}
\sqrt{
\frac{\kappa}{2}
-
\sin^2\frac{\theta}{2}
} & (T\ge T_2, |\theta|<2\arcsin\sqrt{\kappa/2})
\end{array}
\right. 
\label{eq:GWW} 
\end{eqnarray}
For $T\ge T_2$, $\rho(\theta)=0$ at $2\arcsin\sqrt{\kappa/2}\le |\theta|\le\pi$. 
The parameter $\kappa$ is 2 at $T=T_2$ and $\infty$ at $T=T_1$. 
At $T\le T_1$, the system is confined. 
If we interpret the transition at $T=T_2$ as separating the partially and completely deconfined phases, 
it is consistent with the relation \eqref{eq:partial-deconfinement}, just as in Sec.~\ref{sec:4dSYM_S3}. 
\subsection{4d pure Yang-Mills on flat space}
\hspace{0.51cm}
Deconfinement transition in 4d pure Yang-Mills on flat space has been studied extensively on lattice. 
With a lattice regularization, the need for the renormalization makes the determination of the phase distribution tricky, 
namely the `bare' phases observed in the simulation is not necessarily physical; see e.~g.~\cite{Hidaka:2009xh}.  
In Refs.~\cite{Gupta:2007ax,Mykkanen:2012ri}, 4d pure SU($N$) Yang-Mills theory (from $N=3$ to $N=6$) has been studied numerically on lattice, 
and the renormalization has been performed for the Polyakov loop expectation value $\langle P\rangle=\int d\theta e^{i\theta}\rho(\theta)$, so that the zero-point energy is zero. 
At $T=T_2$, the value is close to $\frac{1}{2}$, which is consistent with the GWW-ansatz described above.
(Regarding this obaservation, see Ref.~\cite{Dumitru:2004gd,Nishimura:2017crr} for the argument closely related to ours.)
\subsection{Matrix quantum mechanics}\label{sec:BMN}
\hspace{0.51cm}
Matrix quantum mechanics (i.e. (0+1)-dimensional Yang-Mills) is ultraviolet finite, 
and is an ideal test bed for numerically studying the Polyakov loop expectation value on the lattice, 
without a subtlety of the renormalization. 
\subsubsection*{From unstable to stable with plane wave deformation}
\hspace{0.51cm}
Let us consider the bosonic analogue of the D0-brane matrix model \cite{deWit:1988wri,Witten:1995im,Banks:1996vh,Itzhaki:1998dd}, 
whose Lagrangian is given by 
\begin{eqnarray}
L
=
N{\rm Tr}\left(
\frac{1}{2}\sum_{I=1}^9D_t X_I^2
+
\frac{1}{4}\sum_{I,J=1}^9[X_I,X_J]^2
\right). 
\label{BFSS}
\end{eqnarray}
Here $X_I$ ($I=1,2,,\cdots,9$) are $N\times N$ Hermitian matrices and $D_t X_I = \partial_tX_I-i[A_t,X_I]$ is the covariant derivative of $X_I$.  
This model does not exhibit a first order transition; rather, like the right panel of Fig.~\ref{fig:bosonic-BMN}, 
the model possesses a phase transition at $T_1$ (between the blue and orange lines) and another at $T_2$ (between the orange and red lines).\footnote{
 Note added in ver.~5: 
Recent study \cite{Bergner:2019rca} suggests the transition is actually of first order (see also Ref.~\cite{Azuma:2014cfa}). 
This does not affect our argument substantially; 
depending on the actual order of the phase transition, one of the three patterns discussed in this paper should be realized. 
}
The Polyakov line phases can be numerically well-fit by \eqref{eq:GWW} \cite{Kawahara:2007fn}.

Next let us consider the bosonic version of the plane wave matrix model \cite{Berenstein:2002jq}, 
\begin{eqnarray}
L
&=&
N{\rm Tr}\Biggl(
\frac{1}{2}\sum_{I=1}^9\left(D_t X_I\right)^2
+
\frac{1}{4}\sum_{I,J=1}^9[X_I,X_J]^2
\nonumber\\
& &
\qquad\quad
-
\frac{\mu^2}{2}\sum_{i=1}^3X_i^2 
-
\frac{\mu^2}{8}\sum_{a=4}^9X_a^2 
-
i\sum_{i,j,k=1}^3\mu\epsilon^{ijk}X_iX_jX_k
\Biggl). 
\end{eqnarray}
At $\mu=0$, this is just the model \eqref{BFSS} we have discussed above. As $\mu$ is turned on, 
the phase transition temperatures $T_1$ and $T_2$ gradually approach to each other, 
and eventually coincide, resembling the middle panel of Fig.~\ref{fig:bosonic-BMN}. 
If $\mu$ is tuned even larger, the hysteresis sets in. 
There, the partial deconfinement phase turns to the unstable saddle. 
Intuitively, the reason that the first order transition emerges at large $\mu$ is that 
the eigenvalues of matrices (the location of D-branes) are pushed toward the origin, the off-diagonal matrix elements 
(open strings) can be excited more easily, and the interaction between the D-branes becomes stronger.

\begin{figure}[htbp]
 \begin{minipage}{0.48\hsize}
 \begin{center}
   \rotatebox{0}{
  \includegraphics[width=60mm]{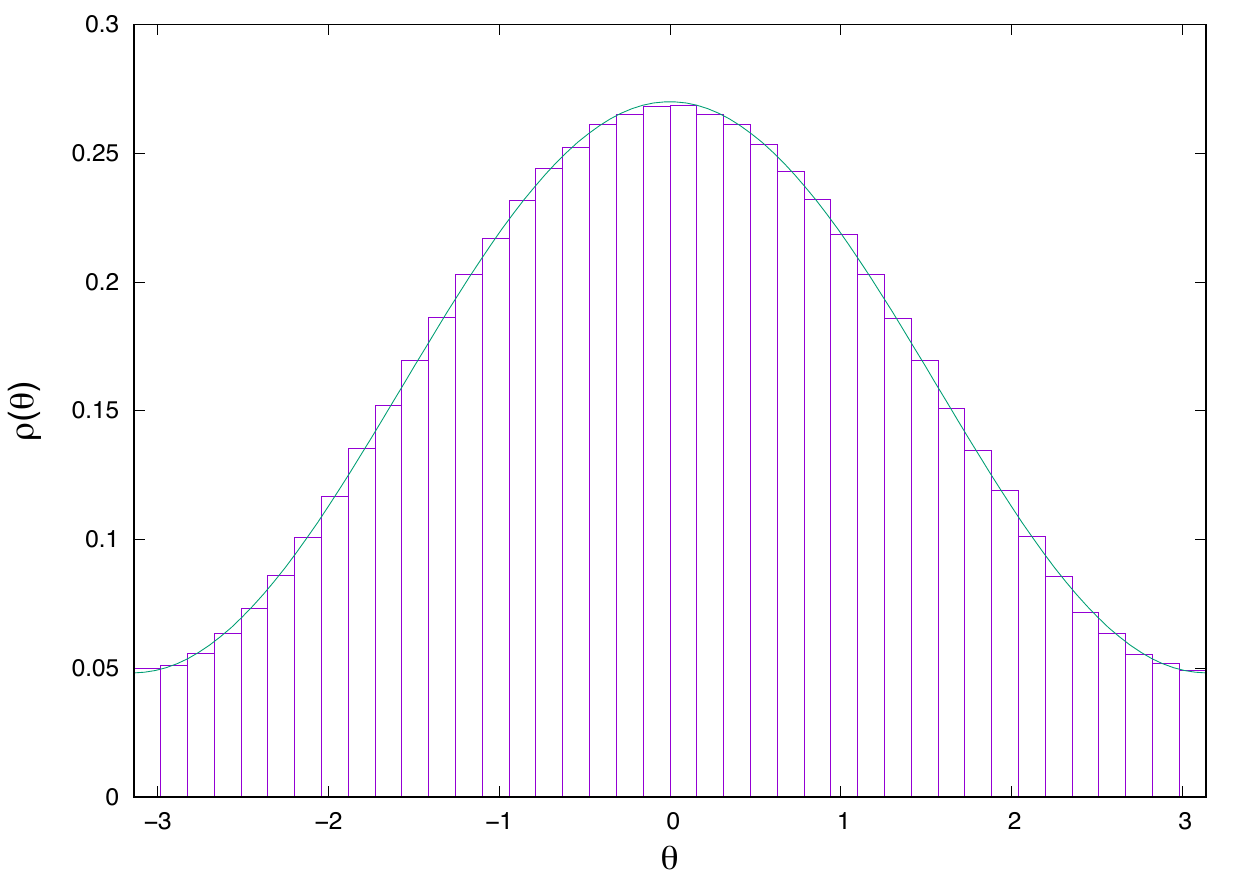}}
 \end{center}
 \end{minipage}
 \begin{minipage}{0.48\hsize}
 \begin{center}
   \rotatebox{0}{
  \includegraphics[width=60mm]{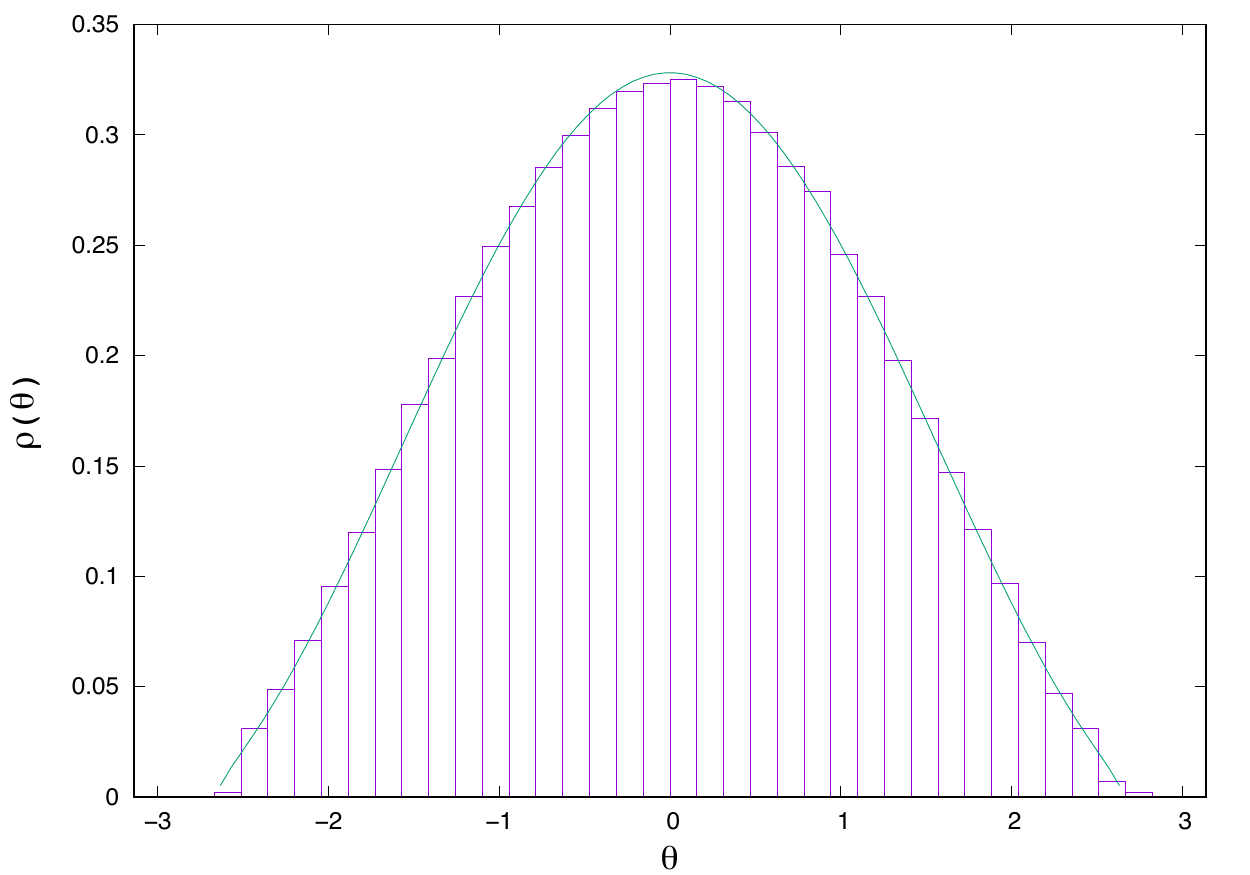}}
 \end{center}
 \end{minipage}
 \caption{
The distribution of Polyakov line phases. 
[Left] $\mu=3.0$, where the partially deconfinement phase is stable. 
We have plotted the distribution at $N=32$, the number of lattice points $L=18$, 
and temperature $T=1.15$, 
which is in the middle of the partially deconfined phase. 
The fit is $\rho(\theta)=\frac{1}{2\pi}\left(
1+\frac{2}{\kappa}\cos\theta
\right)$ with $\kappa=2.87$. 
[Right] $\mu=5.0$, where the partially deconfinement phase is unstable and the transition is of first order. 
We have plotted the distribution at $N=128$, $L=16$, $T=1.54$, which is very close to $T_2$. 
The fit is 
$\rho(\theta)=\frac{2}{\pi\kappa}\cos\frac{\theta}{2}
\sqrt{\frac{\kappa}{2}-\sin^2\frac{\theta}{2}}$ with $\kappa=1.88$. 
 }\label{fig:bosonic-BMN-phase-distribution}
\end{figure}

At small $\mu$ region, we can numerically show that the GWW ansatz \eqref{eq:GWW}  works, 
just in the same way as $\mu=0$; see the left panel of Fig.~\ref{fig:bosonic-BMN-phase-distribution}. 
At $\mu\sim 5$ the first order transition sets in. 
By using the lattice regularization explained in Appendix~\ref{appendix:MM-simulation}, 
with SU(128) gauge group and 16 lattice sites, the hysteresis can be seen around $T=1.55$. 
As expected, in the deconfinement phase $\rho(\theta)$ is consistent with
the last row of \eqref{eq:GWW}, and toward the endpoint ($T=T_2$) 
the value of $\kappa$ approaches $2$ and the gap closes;
see the right panel of Fig.~\ref{fig:bosonic-BMN-phase-distribution}. 

\subsubsection*{D0-brane quantum mechanics}
\hspace{0.51cm}
Now we consider the full, supersymmetric D0-brane matrix model \cite{deWit:1988wri,Witten:1995im,Banks:1996vh,Itzhaki:1998dd}.\footnote{This part is based on forthcoming work with the Monte Carlo String/M-theory Collaboration. The detail will appear in near future, as a part of a separate publication.} 
With the plane wave deformation ($\mu>0$) \cite{Berenstein:2002jq}, this theory is expected to have a first order confinement/deconfinement transition (see e.~g. Ref.~\cite{Costa:2014wya}). 
At $\mu=0$, there are two possibilities: 
(i) the first order transition survives all the way down to $\mu=0$, or 
(ii) it disappears at finite $\mu$. 

In the case of (i), the transition temperature should be zero at $\mu=0$, 
and the confinement phase should disappear, as in the flat space limit of 4d ${\cal N}=4$ SYM. 
Then it is natural to expect the Hagedorn growth exactly at $T=0$, as depicted in the center panel of Fig.~\ref{fig:bosonic-BMN}, 
with $T_1=T_2=0$. 
On the other hand, in Ref.~\cite{Maldacena:2018vsr} it has been argued that the Polyakov line phases
should be almost uniformly distributed and behave like \eqref{eq:phase_partial_deconfinement}.
Then the option (ii), namely the right panel of Fig.~\ref{fig:bosonic-BMN} with $T_1=0$ and $T_2>0$, is the natural candidate.\footnote{
Ref.~\cite{Costa:2014wya} also discussed the possibility that the first order transition disappears at small but finite $\mu$. 
} 

To explore the above possibilities, 
we can use the simulation data obtained from the previous \cite{Berkowitz:2016jlq,Berkowitz:2018qhn}
and ongoing projects by the Monte Carlo String/M-theory Collaboration. Unfortunately, 
this data only covers a moderately high temperature region for which $\langle|P|\rangle\gtrsim 0.7$, 
and hence we cannot judge which possibility is correct. However, we did confirm that the GWW-ansatz holds for the high temperature phase ($T\ge T_2$).

\subsection{2d maximal SYM and black hole/black string topology change}\label{sec:BH-BS}
\hspace{0.51cm}
With the Euclidean signature, the deconfinement transition is the breakdown of the center symmetry along the temporal circle. 
Mathematically, it is the same as the center symmetry along the spatial circle, 
when (one or more of) the spatial dimensions are compactified.\footnote{
The temporal circle has a periodic boundary condition for bosons and an anti-periodic boundary condition for fermions. 
Below, for the spatial circle, we impose periodic boundary conditions for both bosons and fermions. 
} 
The `spatial deconfinement' --- the breakdown of the center symmetry along the spatial circle --- can also be explained by the framework of the partial deconfinement. 

Consider 2d maximal SYM on spatial circle. It is dual to a system of D1-branes in type IIB string theory \cite{Itzhaki:1998dd}, 
which is equivalent to a system of D0-branes in type IIA string theory via T-duality. 
In the D0-brane picture, there can exist a string-like distribution of D0-branes -- called a ``black string'' -- wrapping around the spatial circle.  There can also be black holes, which correspond to a localized distribution of D0 branes.  The distribution of D0-branes is described by the phase distribution of the spatial Polyakov loop.
Thus, we can relate the phase structure of SYM and the black hole/black string system \cite{Aharony:2004ig}.  

There can be `confined', `partially deconfined' and `completely deconfined' phases, 
which correspond to the uniform black string, non-uniform black string and black hole, respectively.\footnote{
Visually the ant trail would look like black string, but actual correspondence is a little bit counter-intuitive. 
Thick, stable ant trail is black hole while the thin trail is non-uniform black string. 
The disappearance of the ant trail is the formation of the uniform black string.
} 
When the non-uniform black string turns into a black hole, the Polyakov line phase distribution $\rho(\theta)$ develops a gap. 
This is consistent with the partial deconfinement picture. 
 
At low temperatures, the dual gravity description is weakly coupled, and a detailed analysis can be done with numerical relativity. 
There are two stable phases --- black hole and uniform black string --- separated by the first order phase transition, and they are connected by the intermediate phase --- nonuniform black string --- \cite{Aharony:2004ig,Catterall:2010fx,Catterall:2017lub,Dias:2017uyv,Harmark:2003yz}. 
The instability toward the deconfined phase is the gauge theory analogue of the Gregory-Laflamme instability \cite{Gregory:1993vy}. 
Also, as mentioned above, the gap at $\theta=\pm\pi$ appears exactly at the border between the black hole and
non-uniform black string phases. 
These can naturally be explained by the partial deconfinement. 

At high temperatures,\footnote{
 Note added in ver.~5: 
Recent study \cite{Bergner:2019rca} suggests the transition is actually of first order (see also Ref.~\cite{Azuma:2014cfa}). 
This does not affect our argument substantially; 
depending on the actual order of the phase transition, one of the three patterns discussed in this paper should be realized. 
} we can use the one-dimensional bosonic matrix model \eqref{BFSS}
to describe the thermal properties of the system. 
In this regime, all three phases appear as stable saddles, and the phase distribution \cite{Kawahara:2007fn} is consistent with the partial deconfinement, as we have already seen in Sec.~\ref{sec:BMN}.

\section{Conclusion and Discussions}\label{sec:discussion}
\hspace{0.51cm}
In this paper, we have argued that the confined and deconfined phases in gauge theories are connected by 
a {\it partially deconfined} phase, which can be stable or unstable depending on the details of the theory. 
When this phase is unstable, it is the gauge theory counterpart of the small black hole phase in the dual string theory. 
Furthermore, we have argued that partial deconfinement is closely related to the Gross-Witten-Wadia (GWW) transition. In order to understand the mechanism behind the partial deconfinement intuitively, 
we have advocated the similarity between the dynamics of gauge theories and the collective behaviors of ants. 
The quantitative evidence is obtained by comparing the distribution of Polyakov line phases with \eqref{eq:partial-deconfinement}. 

A natural, immediate question is whether the partial deconfinement can be seen in the experiments, 
especially in QCD. 
There are a few subtleties in the application of the notion of the partial deconfinement to real-world QCD. 
Firstly, $N=3$ may not be so large. Also the existence of the matter in the fundamental representation, which explicitly breaks the center symmetry,
may modify the nature of the transition. 
Let us still consider possible implications, keeping these subtleties in mind.  
\begin{itemize}
\item
The QCD phase transition at zero chemical potential appears to be a rapid crossover \cite{Aoki:2006we}, which resembles 
the right panel of Fig.~\ref{fig:bosonic-BMN}. 
By changing the quark mass, the first order transition (the left panel of Fig.~\ref{fig:bosonic-BMN}) can also be realized. 
Hence it is natural to expect that the crossover region contains counterparts of partially and completely deconfined phases. 
The partial deconfinement would lead to a new kind of spectrum which has not been considered so far, 
and hence would be phenomenologically relevant. 

\item
It is widely believed that,  at finite chemical potential, the transition becomes first order. 
(See e.~g. \cite{Muroya:2003qs} for a review.)
If this is true, it is natural to expect that the partially deconfined unstable phase --- the gauge theory counterpart of the Schwarzschild black hole --- exists.
It would be a useful setting for `experimental realization' of evaporating black hole!
Note that the stringy correction should be large
and hence it would not be very close to the black hole in weakly coupled gravity.
Nonetheless, it would be interesting if some universal features could be seen.

\item
A few remarks regarding the nature of the phase transition at finite chemical potential is in order here.\footnote{
We would like to thank K.~Fukushima for useful comments regarding these points. 
} 

Usually it is argued that the chiral transition (characterized by the growth of chiral condensate from $\simeq 0$ to a large value), which can be different from deconfinement (characterized by the growth of the Polyakov loop) in general, becomes first order.
Whether these two transitions are correlated as $\mu=0$ case is not clear because the lattice QCD simulation is difficult 
due to the sign problem.

Whether two transitions coincide or not,
the unstable phase can exist as long as there is a hysteresis. 
At a possible first order chiral transition, 
if the Polyakov loop jumps as well --- whether zero to nonzero, or nonzero to nonzero ---
the argument for the small black hole phase presented in Sec.~\ref{sec:ant-BH-correspondence} would apply
without change.\footnote{
A subtlety here is that the Polyakov loop may not be a good quantity to characterize the partial deconfinement, because of the existence of the fundamental fermion.  
}  
Note that, in the case of 4d ${\cal N}=4$ super Yang-Mills, the small black hole phase connected 
the completely deconfined and Hagedorn string phases, both correspond to $P>0$.

\item
Given that we might be able to see the `evaporating black hole' experimentally, 
it is interesting to study possible experimental signals associated with the first order phase transition. 
The nature of the phase transitions discussed so far looks rather universal. 
Hence it would be useful to consider `applied AdS/CFT' based on 4d SYM on S$^3$ (see e.g.~Ref.~\cite{Jokela:2015sza}). 

\end{itemize}

There are several other issues which are important from the quantum gravity point of view:

--- Analyses of the gravitational dual of 4d maximal SYM suggests the existence of 
a first order transition corresponding to the localization of the black hole along S$^5$, 
even in the microcanonical treatment \cite{Dias:2016eto,Yaffe:2017axl}. 
This cannot be explained by the gauge theory argument in this paper. 
We emphasize that the color degrees of freedom would play a crucial roles in this case as well. 
One possibility is that a GWW-like transition in the microcanonical treatment describes the transition between 
the positive and negative specific heat parameter regions of the AdS$_5$ black hole which are not localized along S$^5$, 
and the transition associated with the localization along S$^5$ exists separately. 
It may well be the case that the localization along S$^5$ (i.e. breaking of the R-symmetry) corresponds to the GWW transition. 
That the previous weak-coupling analyses did not capture this transition would mean that the stringy corrections resolve it. 

--- When Refs.~\cite{Berkowitz:2016znt,Berkowitz:2016muc} discussed the negative specific heat in the context of gauge/gravity duality, the D0-brane matrix model has been considered as a concrete example. 
Instead of the confinement, the Higgs mechanism associated with the emission of the eigenvalues due to the flat direction in the moduli
has been used for the reduction of the unlocked degrees of freedom. 
For the flat direction to exist, supersymmetry is crucial, and hence the generalization of the argument to nonsupersymmetric systems, 
like our universe, was not clear. The partial deconfinement is almost the same (i.e. some degrees of freedom are confined rather than Higgsed) but 
supersymmetry is not required. 

--- The numerical tests explained in Sec.~\ref{sec:polyakov_line} did not directly consider the unstable saddle. 
In principle, we could directly see the unstable saddle of the path integral \eqref{eq:canonical-partition-function}, by counting the number of Monte Carlo samples
for each value of $E$ and determining $F(E,T)$ in \eqref{eq:canonical-partition-function}. 
Of course, a na\"{i}ve simulation samples only the configurations around the minima,  
but we might be able to circumvent this problem by restricting the energy to a narrow band.\footnote{
Essentially the same manipulation for the path integral in the gravity side has been considered in Ref.~\cite{Marolf:2018ldl}. 
}$,$\footnote{
Such procedure is straightforward for the theories without the sign problem.
In many supersymmetric theories, the sign problem does exist, namely the determinant of the Dirac operator is complex, 
and hence we need to find a way to fix the energy taking into account the effects of the complex phase.
Fortunately, the complex phase is often close to 1, so that it can simply be neglected (see ref.~\cite{Schaich:2018mmv} for recent review).   
Even when the complex phase fluctuates violently, it is not correlated to various physical quantities and can be neglected, 
at least in the case of D0-brane matrix model \cite{Hanada:2011fq,Berkowitz:2016jlq}.
Therefore, there is a fair chance that such strategy can work even for supersymmetric theories which have good weakly-coupled gravity duals simply neglecting the phase. 
} 
Whether such a sampling procedure works with reasonable computational resources is not clear at this moment, but it is certainly worth trying.
It is also important to derive the negative specific heat analytically, away from the weak coupling region.  
See Ref.~\cite{Berenstein:2018hpl} for a recent attempt. 

Last but not least, the correspondence between the ant trail and black hole is not perfect. 
The biggest difference is that $N_{\rm trail}$ can reach $N$ only at $p=\infty$, and as a consequence, a GWW-like transition is missing. 
It would be interesting if there is a more precise analogue of black holes which exhibits the GWW transition. 
See an improved model discussed in Appendix~\ref{sec:modified-ant-equation}
(the Ant-Man model) for an attempt along this direction. 

\begin{center}
\section*{Acknowledgements}
\end{center}
\hspace{0.51cm}
We thank Tatsuo Azeyanagi, David Berenstein, Ramy Brustein, Jordan Cotler, Oscar Dias, Pau Figueras, 
Kenji Fukushima, Hrant Gharibyan, Yoshimasa Hidaka, Juan Maldacena, 
Marco Panero, Rob Pisarski, Enrico Rinaldi, Jorge Santos, 
Bo Sundborg, Masaki Tezuka and Naoki Yamamoto
for discussions and comments. 
We thank Jordan Cotler also for carefully reading the manuscript many times and providing us with countlessly many comments.  
M.~H. thanks Brown University for the hospitality during his stay while completing the paper, 
and acknowledges the STFC  Ernest  Rutherford  Grant ST/R003599/1.  
M.~H. and G.~I. acknowledge JSPS KAKENHI Grants 17K14285 and 16K17679, respectively.
The work of G. I. was also supported, in part, by Program to Disseminate Tenure Tracking System, MEXT, Japan.

We originally learned about the ant model from Ref.~\cite{Sumpter:2016}. 
We thank British people's devotion to the beautiful game of football, which made this kind of book commercially successful.

\appendix
\section{Small black hole from gauge theory}\label{sec:Hanada-Maltz}
\hspace{0.51cm}
In this Appendix, we review the heuristic explanation of the black hole equation of state
from the dual 4d super Yang-Mills. We closely follow Ref.~\cite{Hanada:2016pwv}. 
The most basic assumption of the partial deconfinement introduced in Ref.~\cite{Hanada:2016pwv} is that,  
regardless of the value of $N_{\rm BH}$, the property of the small black hole (unstable saddle) is the same as the one at $T=T_2$ and $N_{\rm BH}=N$. 
Let $X_{\rm BH}$ be the $N_{\rm BH}\times N_{\rm BH}$ sub-matrix of the scalar $X$, which corresponds to the partially deconfined SU$(N_{\rm BH})$. 
As $N_{\rm BH}$ changes, the 't Hooft coupling effectively changes as
$g_{\rm YM}^2N_{\rm BH}=(g_{\rm YM}^2N)\cdot (N_{\rm BH}/N)=\lambda\cdot (N_{\rm BH}/N)\equiv \lambda_{\rm BH}$. 

When $\lambda_{\rm BH}\ll 1$, the $N_{\rm BH}\times N_{\rm BH}$ block is weakly coupled, and hence we simply see the Hagedorn growth. 
On the other hand, 
when $\lambda_{\rm BH}\gg 1$, the $N_{\rm BH}\times N_{\rm BH}$ block is strongly coupled. 
Then the potential term $\frac{1}{g_{\rm YM}^2}{\rm Tr}[X_{{\rm BH}I},X_{{\rm BH}J}]^2
=\frac{N_{\rm BH}}{\lambda_{\rm BH}}{\rm Tr}[X_{{\rm BH}I},X_{{\rm BH}J}]^2$ is more important than the kinetic term of the action.
By using $Y_{\rm BH}\equiv X_{\rm BH}/\lambda_{\rm BH}^{1/4}$, the potential becomes independent of the coupling constant. 
As mentioned in Sec.~\ref{sec:polyakov_line}, we identify the temperature $T$ to be the lowest possible temperature 
in the SU($N_{\rm BH}$) theory at which all $N_{\rm BH}$ D-branes can form the bound state. 
Given that there is no particular $\lambda_{\rm BH}$ dependence in terms of $Y_{\rm BH}$, 
it is natural to suppose that the eigenvalues of $Y_{\rm BH}$ at $T$ are independent of $\lambda_{\rm BH}$, 
and hence the eigenvalues of $X_{\rm BH}$ scale as $\lambda_{\rm BH}^{1/4}$. 
Intuitively, the eigenvalue of $X_{\rm BH}$ corresponds to the radius of the black hole, which sets a natural inverse energy scale. 
So it is natural to expect $T\sim\lambda_{\rm BH}^{-1/4}\sim (N_{\rm BH}/N)^{-1/4}$.\footnote{
Intuitively, at higher temperatures more open strings are excited, D-branes are more tightly bound, 
and the radius of the black hole becomes smaller. 
This is the same as the scaling of $\ell_s$ in the dual frame ($R_{{\rm S}^3}$ and $R_{\rm AdS}$ fixed). 
}  
Also, from the dimensional analysis and 't Hooft counting, the energy and entropy should scale as 
$E\sim N_{\rm BH}^2T\sim N^2 (N_{\rm BH}/N)^{7/4}\sim N^2T^{-7}$
and $S\sim N_{\rm BH}^2\sim N^2 (N_{\rm BH}/N)^2\sim N^2T^{-8}$. 
(The 't Hooft counting alone cannot fix the dependence on the 't Hooft coupling.
Here we implicitly assumed that it is independent of the coupling in the strongly coupled region.
This is equivalent to assuming that the Newton constant is independent of $\lambda$ in the dual frame ($R_{{\rm S}^3}$ and $R_{\rm AdS}$ fixed), 
which can be confirmed analytically in the supersymmetry-preserving setups.)

Next let us consider the M-theory parameter region of the ABJM theory \cite{Aharony:2008ug}.\footnote{
Whether the same argument holds in the type IIA string theory limit (typically $N/k$ fixed) is a subtle issue, 
because the moduli $(\mathbb{C}^4/\mathbb{Z}_k)^N/({\rm permutation})$, and the dual gravity geometry AdS$_4\times$S$^7/\mathbb{Z}_k$, 
explicitly depend on $k\sim N$. If we apply the same power counting na\"{i}vely, we obtain $E\sim \frac{1}{G_{\rm N,10}T^{-8}}$, 
which is not the equation of state of the ten dimensional black hole.   
} 
We take the Chern-Simons level $k$ to be $O(1)$, and send $N$ to infinity. 
The potential term of the action is schematically $\sim \frac{N}{\lambda}{\rm Tr}X^6$ ($\lambda=\frac{N}{k}$), where $X$ stands for the scalar fields in the bifundamental representation. 
The $\lambda$ dependence disappears if we use $Y\equiv\lambda^{-1/6}X$, 
and hence the natural scaling of the eigenvalues is $\lambda_{\rm BH}^{1/6}$. 
This motivates $T\sim (N_{\rm BH}/N)^{-1/6}$. Note that this is the scaling of the Planck scale in the dual frame. 
The Newton constant is $G_{\rm N,11}\sim\sqrt{\lambda}/N^2\sim N^{-3/2}$,  
which can be checked by supersymmetric localization \cite{Pestun:2016zxk}. 
Hence the entropy is $S\sim N_{\rm BH}^{3/2}\sim N^{3/2}T^{-9}\sim \frac{1}{G_{\rm N,11}T^{9}}$, 
and the energy is $E\sim \frac{1}{G_{\rm N,11}T^{8}}$. 
This is the correct behavior in the eleven dimensional supergravity. 

\section{The large $N$ limit of the ant equation}\label{sec:ant-large-N}
\hspace{0.51cm}
In Ref.~\cite{Beekman9703}, the number of the ants in the colony $N$ is varied 
while other parameters $\alpha$, $s$ and $p$ are fixed. 
In order to see the resemblance between the ant model and the holographic description of a black hole, 
it is convenient to fix $N$ to be a large value. Like in string theory and quantum field theory, 
we expect mathematical simplifications to arise in the natural large-$N$ (many-ant) limit. 
Furthermore, in the large-$N$ limit, the phase transition appears in the mathematically precise sense. 

Let us divide the left and right hand sides of \eqref{eq:ant-equation} by $N$. 
By using $x\equiv N_{\rm trail}/N$, $\tilde{p}\equiv pN$ and $\tilde{s}\equiv s/N$, 
we obtain 
\begin{eqnarray}
\frac{dx}{dt}
=
(\alpha+\tilde{p}x)(1-x)
-
\frac{\tilde{s}x}{\tilde{s}+x}.
\end{eqnarray}
We treat this $x$ as an order $N^0$ quantity. 
The `ant equation' $\frac{dx}{dt}=0$ becomes
\begin{eqnarray}
f(x)\equiv \tilde{p}x^3
+
\left(
\alpha-\tilde{p}+\tilde{p}\tilde{s}
\right)x^2
+
\left(
\alpha\tilde{s}-\tilde{p}\tilde{s}-\alpha+\tilde{s}
\right)x
-
\alpha\tilde{s}
=0.
\label{eq:ant-equation-2} 
\end{eqnarray}
By definition, 
\begin{eqnarray}
0\le x\le 1
\end{eqnarray}
is required. 
Also, by `physical' reasoning, we should impose 
\begin{eqnarray}
\alpha, \tilde{p},\tilde{s}\ge 0. 
\end{eqnarray}
Here $p=\tilde{p}/N$ is the strength of the pheromone secreted by each ant. 
In the 't Hooft large-$N$ limit of the gauge theory, the 't Hooft coupling $g_{\rm YM}^2N$ is fixed, so that the interaction does not become too strong. The analogous scaling is $\alpha,p\sim N^{-1}$.\footnote{For the actual ants, 
$\alpha,p\sim N^0$ would be more natural. 
This corresponds to $g_{\rm YM}^2$ fixed, which is analogous to M-theory. 
}
The value of $p$ can change depending on the environment around the nest. 
For example, if the air is dry, the pheromone can evaporate quickly and hence $p$ is smaller. 

The parameter $s$ controls the rate that ants leave the trail; maybe they get bored, maybe they get tired. 
It is natural to assume that the number of such ants is proportional to $N$ and hence $\tilde{s}$ is of order one.

For fixed $\alpha N$ and $\tilde{s}$, we solve \eqref{eq:ant-equation-2} for various values of $\tilde{p}$
and plot $x$ as a function of $\tilde{p}$. 
Two limiting situations, $\tilde{p}\to 0$ and $\tilde{p}\to\infty$, are easy:
\begin{itemize}
\item
When $\tilde{p}\to 0$, 
$\alpha x^2
+
\left(
\alpha\tilde{s}-\alpha+\tilde{s}
\right)x
-
\alpha\tilde{s}
=0$, and hence $x\simeq\alpha$, 
namely $x$ is almost zero.
In words: if there is no pheromone, there is no trail. 

At finite $\tilde{p}$, the solution close to zero can be written as $x=\frac{\alpha }{1-\tilde{p}}+O(\alpha ^2)$. 
The deviation from zero becomes large when $1-\tilde{p}$ becomes of order of $\alpha $. 

\item
When $\tilde{p}\to\infty$, 
$\tilde{p}x^3
+
\left(
-\tilde{p}+\tilde{p}\tilde{s}
\right)x^2
-\tilde{p}\tilde{s}x
$
has to be of order one, and hence 
$x^3
+
\left(
-1+\tilde{s}
\right)x^2
-\tilde{s}x
$
has to be almost zero. 
Then $x\to 1$.
($x\to 0$ may appear to be fine as well, but it requires $x\simeq -\frac{\alpha }{\tilde{p}}<0$ and hence is `unphysical'.
Yet another solution $x\simeq-\tilde{s}$ is also `unphysical'.) 
Namely, if the pheromone is extremely strong, all ants join the trail. 

Let us write $x=1-\epsilon$ there. Then, $\epsilon\simeq\frac{\tilde{s}}{\tilde{p}(\tilde{s}+1)}$. 
\end{itemize}

The deviation from `confinement' $x\sim\alpha $ sets in at $\tilde{p}\simeq 1$. 
If there is another solution to the ant equation there, we should see an $S$-shape curve (i.e. the first order transition). 
At $\tilde{p}=1$, the ant equation is $x^3-(1-\tilde{s})x^2\simeq 0$. 
Hence, if $\tilde{s}<1$, there are two solutions near $x=0$ (analogous to `confined' and `partially deconfined' phases)
and the other near $x=1-\tilde{s}$ (analogous to a `completely deconfined' phase).

Let us study the shape of the $S$-curve more precisely. 
It is described by 
\begin{eqnarray}
\tilde{p}x^3
-
\tilde{p}(1-\tilde{s})x^2
+
(1-\tilde{p})\tilde{s}x
\simeq
0. 
\end{eqnarray}
The `partially' and `completely' deconfined phases are 
\begin{eqnarray}
x\simeq \frac{1-\tilde{s}\pm\sqrt{(1-\tilde{s})^2-\frac{4(1-\tilde{p})\tilde{s}}{\tilde{p}}}}{2}. 
\label{saddles}
\end{eqnarray}
The bending point is where $\sqrt{(1-\tilde{s})^2-\frac{4(1-\tilde{p})\tilde{s}}{\tilde{p}}}$ becomes zero, namely 
\begin{eqnarray}
\tilde{p}=\frac{4\tilde{s}}{(1-\tilde{s})^2+4\tilde{s}}. 
\end{eqnarray}

As $\tilde{s}$ approaches zero, the bending point gets closer to 0 and the hysteresis becomes stronger. 
This behavior is `physically' almost trivial: if no ant leaves the trail, 
no pheromone is needed to keep the existing trail. 
The small $\tilde{s}$ region is similar to the strong coupling region of 4d SYM.

As $\tilde{s}$ approaches 1, the hysteresis becomes weaker, and the unstable saddle becomes closer to the vertical straight line near $x=0$.  
This resembles the Hagedorn string. 

\begin{figure}[htbp]
 \begin{minipage}{0.5\hsize}
  \begin{center}
   \scalebox{2}{
  \includegraphics[width=50mm]{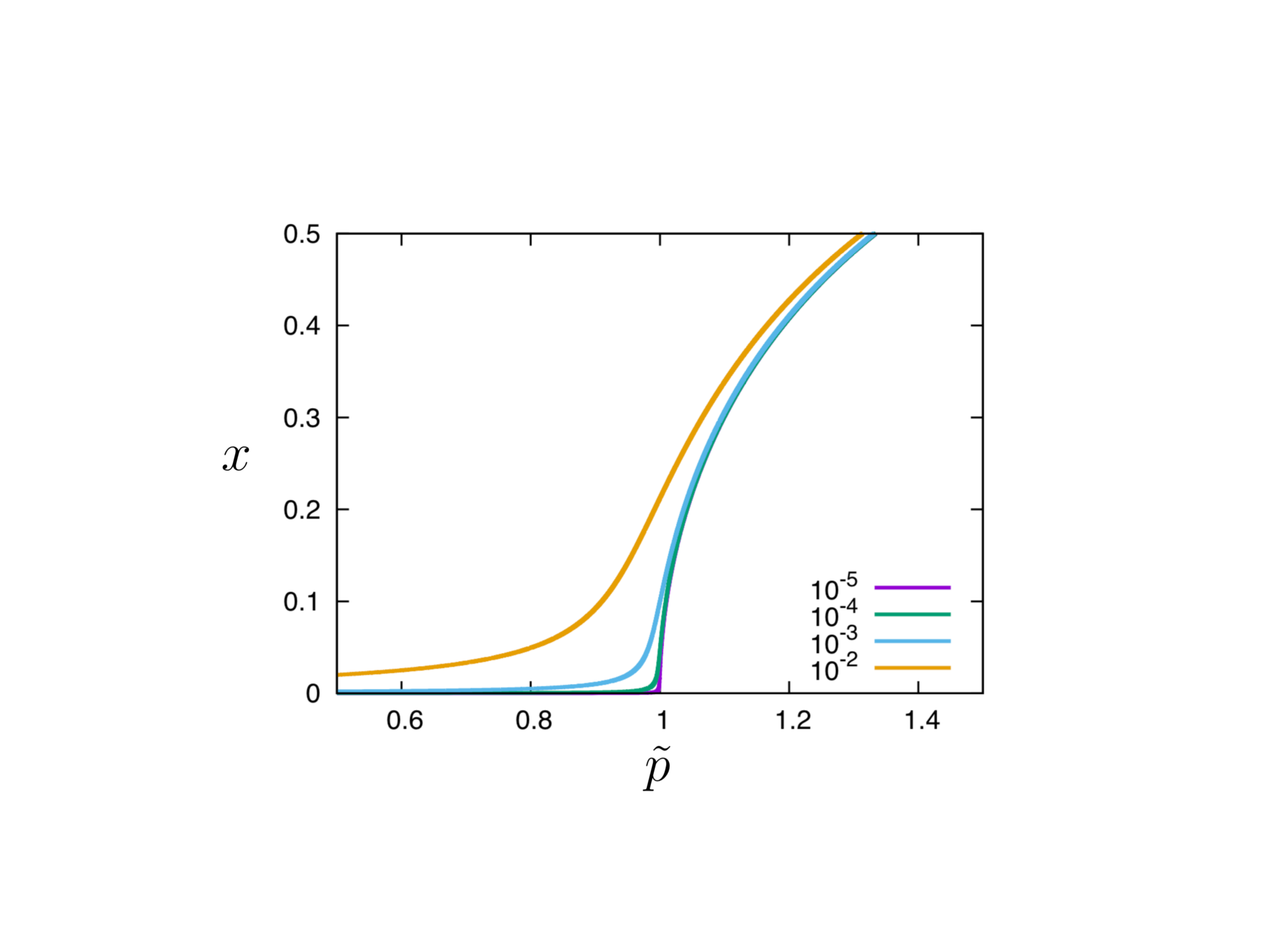}}
 \end{center}
 \end{minipage}
 \begin{minipage}{0.5\hsize}
 \begin{center}
   \scalebox{2}{
  \includegraphics[width=50mm]{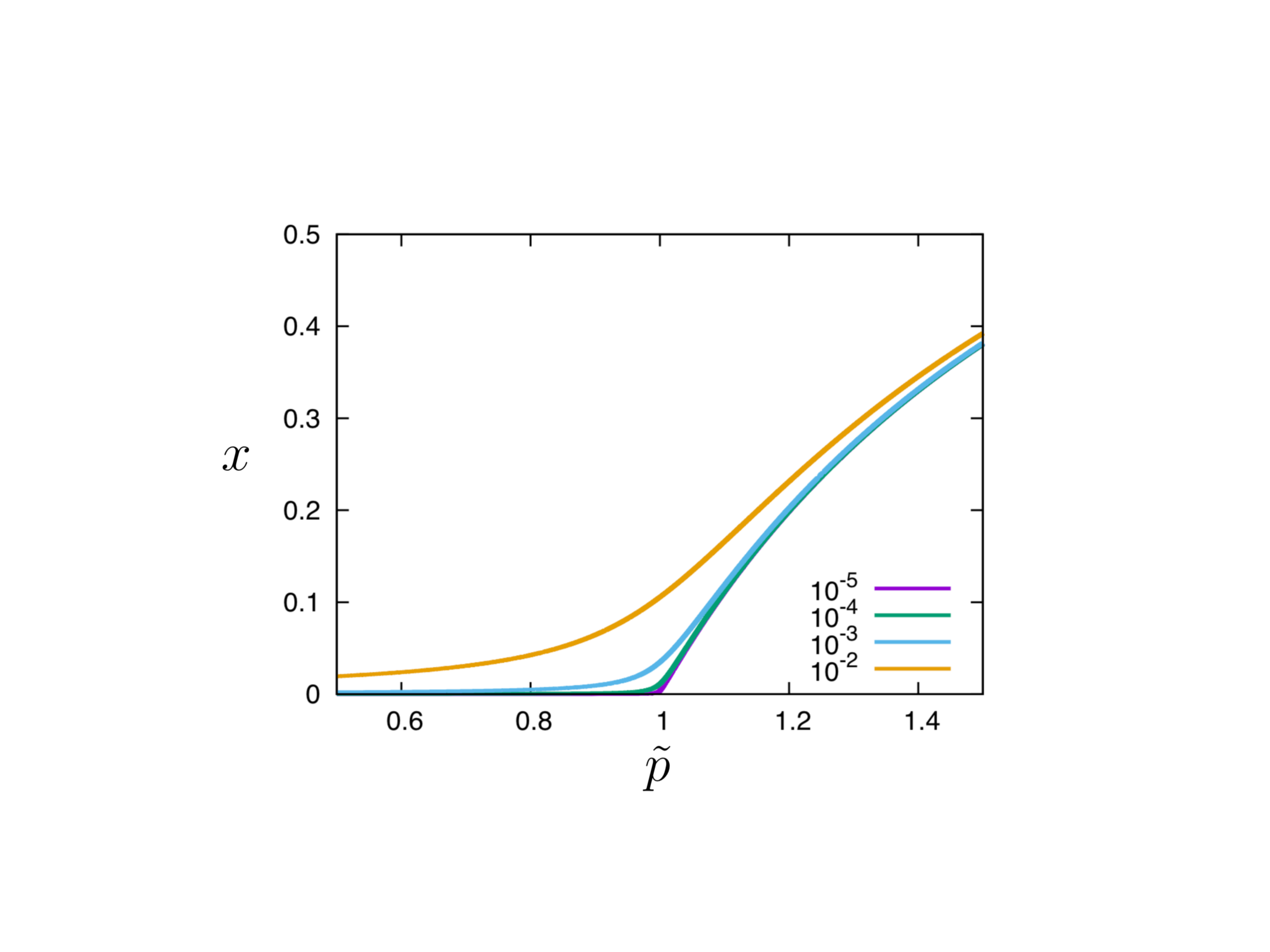}}
 \end{center}
 \end{minipage}
 \caption{$x$ vs $p$ with $\tilde{s}=1$ (left) and $\tilde{s}=5$ (right), $\alpha =10^{-2}, 10^{-3}, 10^{-4}$ and $10^{-5}$. 
 We can see the emergence of the singularity at $\tilde{p}=1$ as $\alpha $ approaches zero, or equivalently, in the large-$N$ limit. 
 }\label{fig:2nd-order-transition}
\end{figure}

For $\tilde{s}>1$, the `partially deconfined' phase becomes stable.
This is analogous to the cross-over in real-world QCD. 
Right above $\tilde{p}=1$, one of \eqref{saddles} describes this phase. By writing $\tilde{p}=1+q$, we obtain 
\begin{eqnarray}
x\simeq\frac{q\tilde{s}}{\tilde{s}-1}. 
\end{eqnarray}
As $q\to 0$, we have that $x$ is continuous, but $\frac{dx}{dq}$ is not. 
Thus the transition from `confinement' to `partial deconfinement' is of second order in this case, 
just like the case of large-$N$ Yang-Mills.  
Note that the large-$N$ limit, which is analogous to the thermodynamic limit in statistical physics, is necessary for the phase transition to take place; see Fig.~\ref{fig:2nd-order-transition}. 

Although the transition from `confinement' to `partial deconfinement' is nicely captured by the ant model, 
the transition to the `complete deconfinement' is not very precisely captured. 
At $\tilde{s}\to 0$, the value of $x$ at the bending point is $1/2$, which is different from the value we expect in SYM, 1. 
At $\tilde{s}=1$, the analogue of the Hagedorn growth does not continue up to $x=1$. 
As we have already seen, the large-$\tilde{p}$ asymptotic behavior is 
$x\simeq 1-\frac{\tilde{s}}{\tilde{p}(\tilde{s}+1)}$. Hence the `complete deconfinement' is achieved only at $\tilde{p}=\infty$. 
The same applies to $\tilde{s}>1$ as well. Hence the GWW transition, 
which separates the partially and completely deconfined phases, is not captured by the ant model. 
It would be intereting if we can find other complex systems which mimic black holes more precisely. 
Below, we give a toy model which captures the complete deconfinement.

\subsection{Modified ant model (Ant-Man model)}\label{sec:modified-ant-equation}
\begin{figure}[htbp]
 \begin{minipage}{0.32\hsize}
 \begin{center}
   \scalebox{1.3}{
  \includegraphics[width=50mm]{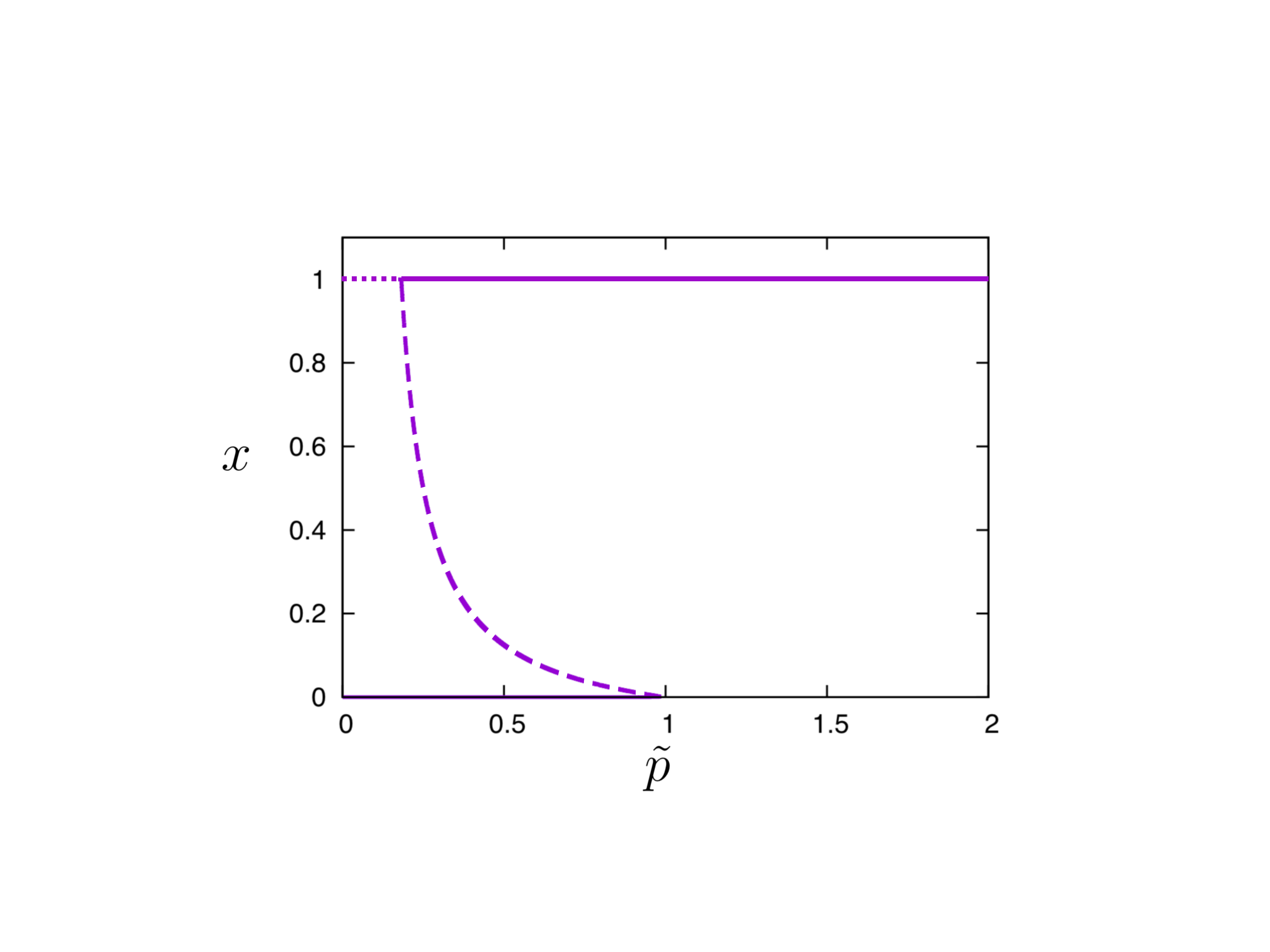}}
 \end{center}
 \end{minipage}
 \begin{minipage}{0.32\hsize}
 \begin{center}
   \scalebox{1.3}{
  \includegraphics[width=50mm]{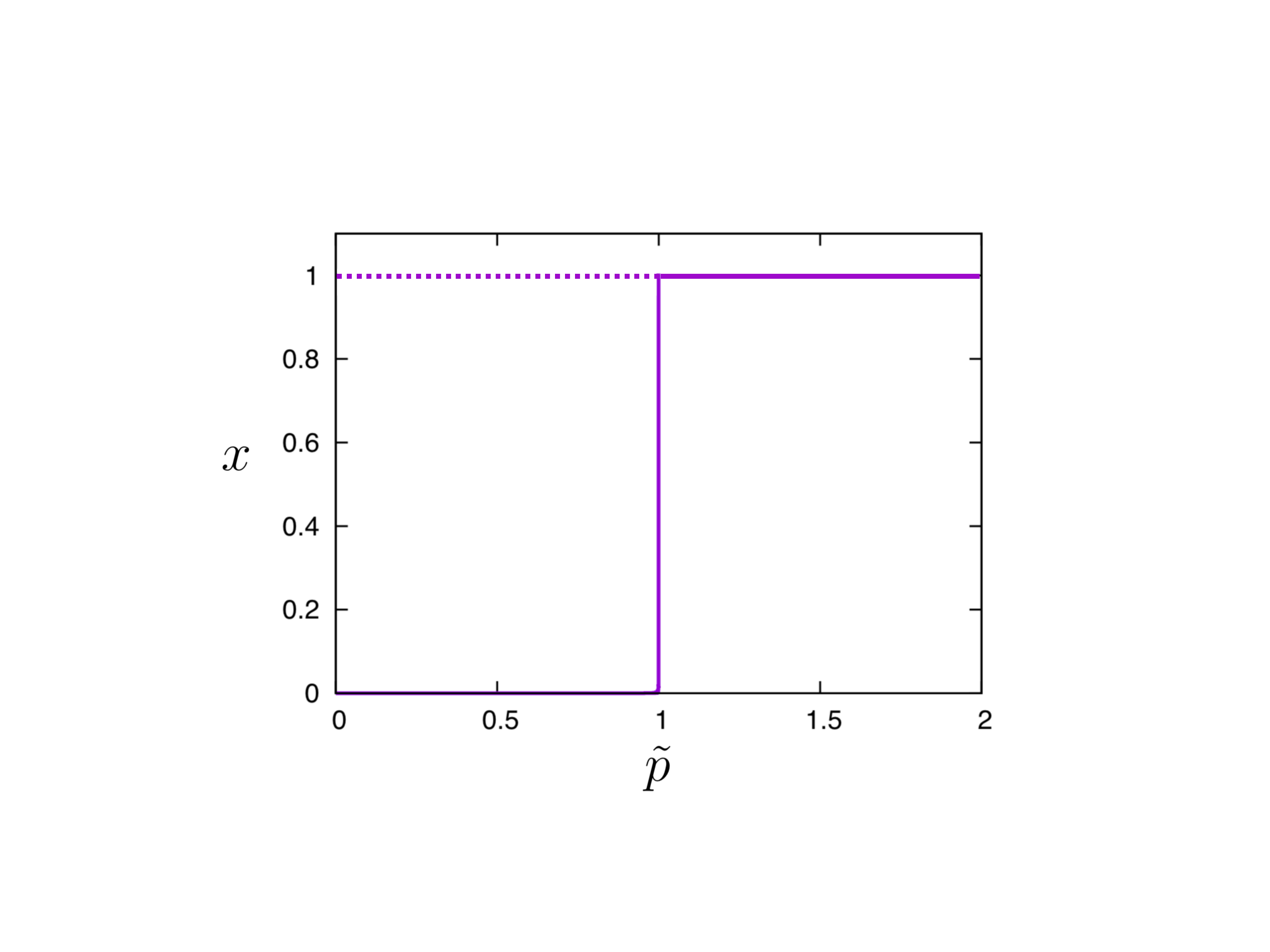}}
 \end{center}
 \end{minipage}
  \begin{minipage}{0.32\hsize}
 \begin{center}
   \scalebox{1.3}{
  \includegraphics[width=50mm]{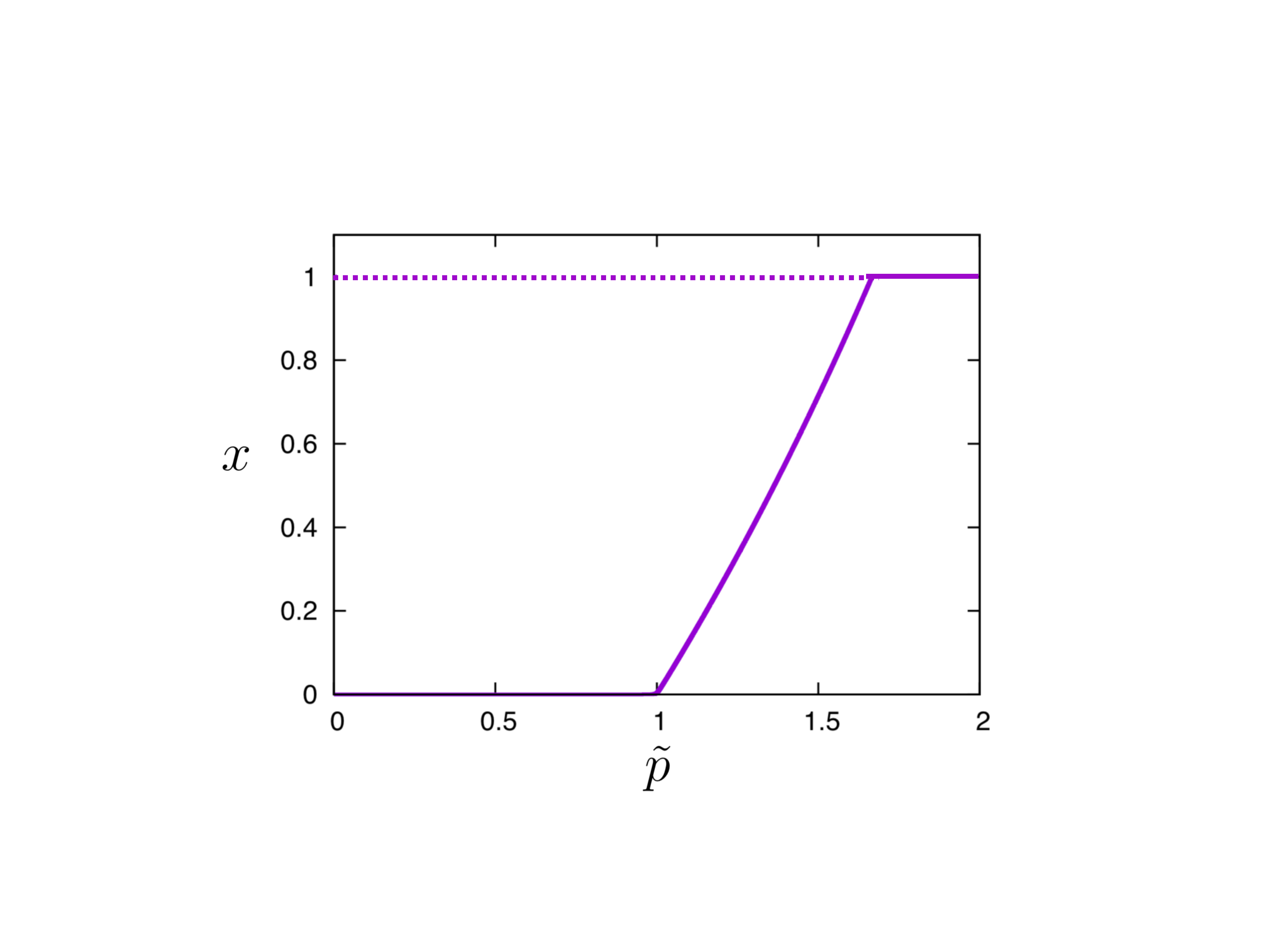}}
 \end{center}
 \end{minipage}
\caption{
$x\equiv\frac{N_{\rm trail}}{N}$ versus $\tilde{p}$ in the modified ant model \eqref{eq:modified-ant-equation},  
with $\alpha=\frac{1}{N}$, $\tilde{s}\equiv\frac{s}N=0.1$ (left), $1.0$ (middle) and $5.0$ (right), 
$N=10^5$.  
 }\label{fig:modified-ant-equation}
\end{figure}

\hspace{0.51cm}
Here we consider a modified ant model --- `Ant-Man model' --- which gives the ants some human characteristics.
We assume they think like us humans: 
when everybody is on the trail, they are scared to leave. 
If such human-ish ants existed, the ant trail model would be modified as 
\begin{eqnarray}
\frac{dx}{dt}
=
(\alpha +\tilde{p}x)(1-x)
-
\frac{\tilde{s}x}{\tilde{s}+x}\cdot (1-x^2).
\label{eq:modified-ant-equation}
\end{eqnarray}
Here the factor $1-x^2$ multiplies to the outflow factor, 
so that no ant leaves the trail when all ants are on the trail ($x=1$).  
Taking into account the constraint $0\le x\le 1$, the phase structures become those in Fig.~\ref{fig:modified-ant-equation}. 
They are closer to Yang-Mills (Fig.~\ref{fig:Nbh-vs-T}), except that new unstable trail ($x=1$, shown by the dotted line) appears.
This unwanted unstable trail is actually harmless; 
it can be eliminated by a small change such as $(1-x^2)\to (1+\epsilon-x^2)$ with infinitesimally small positive $\epsilon$ (i.e. a little bit of courage to leave the trail).  

We emphasize that we have written down this model just to mimic Yang-Mills more closely. 
There is no theoretical or experimental justification of the existence of Ant-Men. 
Still, such models would be useful for obtaining the intuition into the nature of Yang-Mills and string theory.

\section{Monte Carlo simulation of the matrix model}\label{appendix:MM-simulation}
\hspace{0.51cm}
The simulation we have used for the study of the bosonic matrix model is based on 
the simulation code for the BMN matrix model written by M.~H. 
for the Monte Carlo String/M-theory collaboration.  
We have just removed fermions from the code.\footnote{The code is available upon request to M.~H.} 

The lattice action is 
\begin{eqnarray}
S_{\rm Lattice}
&= &
\frac{N}{2a}\sum_{t,M}{\rm Tr}\left(DX_M(t)\right)^2
-
\frac{Na}{4}\sum_{t,M,N}{\rm Tr}[X_M(t),X_N(t)]^2
\nonumber\\
& &
+
aN\sum_t\ {\rm Tr}\left\{
\frac{\mu^2}{2}\sum_{i=1}^3X_i(t)^2 
+
\frac{\mu^2}{8}\sum_{a=4}^9X_a(t)^2 
+
i\sum_{i,j,k=1}^3\mu\epsilon^{ijk}X_i(t)X_j(t)X_k(t)
\right\}
\nonumber\\
& &
-
\sum_{i<j}2\log\left|\sin\left(\frac{\alpha_i-\alpha_j}{2}\right)\right|, 
\end{eqnarray}
where $U={\rm diag}(e^{i\alpha_1/N_t},e^{i\alpha_2/N_t}\cdots,e^{i\alpha_N/N_t})$,  
$-\pi\le \alpha_i<\pi$, and 

\begin{eqnarray}
DX(t)
\equiv
\frac{1}{2}U^2 X(t\pm 2a)\left(U^\dagger\right)^2
+ 2U X(t\pm a)U^\dagger
-\frac{3}{2} X(t)
=
aD_t X(t) + O(a^3). 
\end{eqnarray} 
We simulate this action by using the Hybrid Monte Carlo algorithm. 

The Polyakov loop is defined by 
\begin{eqnarray}
P=\frac{1}{N}\sum_{j=1}^Ne^{i\alpha_j}. 
\end{eqnarray}
The phases are $\alpha_1,\alpha_2,\cdots,\alpha_N$. 
We need to take into account the ambiguity of the global U$(1)$ factor, with which all $\alpha$'s are shifted by a constant. 
We fix it by requiring $P=|P|$. 

We have studied $\mu=1,2,3$ and $4$ with matrix size $N=32$, lattice size $L=18$. 
In this region, we obtained the data consistent with the existence of stable partially deconfined phase, 
although a first order transition may emerge at larger $N$. 
We have also studied $\mu=5$ with $N=128$ and $L=16$. There, we have observed a clear hysteresis
around $T=1.55$; namely, two different phases are observed with hot start (lower the temperature gradually)
and cold start (raise the temperature gradually).

\bibliographystyle{utphys}
\bibliography{small-BH}

\end{document}